\documentclass[sigconf, 9pt]{acmart}
\usepackage{amsfonts}
\usepackage[linesnumbered,algoruled]{algorithm2e}
\usepackage{algpseudocode}
\usepackage{amsthm}
\usepackage{graphicx}
\usepackage{textcomp}
\usepackage[font={small, it},skip=1pt]{caption}

\usepackage{subcaption}
\usepackage{xcolor}
\usepackage{booktabs}
\usepackage{todonotes}
\SetKwInOut{Parameter}{parameter}
\theoremstyle{theorem}
\newtheorem{example}{Example}[section]

\expandafter\def\expandafter\example\expandafter{\example\small}

\SetKwRepeat{Do}{do}{while}

\AtBeginDocument{%
	\providecommand\BibTeX{{%
			\normalfont B\kern-0.5em{\scshape i\kern-0.25em b}\kern-0.8em\TeX}}}

\setcopyright{acmcopyright}
\copyrightyear{2020} 
\acmYear{2020} 
\setcopyright{acmcopyright}\acmConference[ICCAD '20]{IEEE/ACM International Conference on Computer-Aided Design}{November 2--5, 2020}{Virtual Event, USA}
\acmBooktitle{IEEE/ACM International Conference on Computer-Aided Design (ICCAD '20), November 2--5, 2020, Virtual Event, USA}
\acmPrice{15.00}
\acmDOI{10.1145/3400302.3415681}
\acmISBN{978-1-4503-8026-3/20/11}

\begin{document}
	
\title{CONTRA: Area-Constrained Technology Mapping Framework For Memristive Memory Processing Unit}

	
	\author{Debjyoti Bhattacharjee}
	\orcid{1234-5678-9012}
	\affiliation{%
		\institution{IMEC, Leuven, Belgium}
	}
	\email{debjyoti.bhattacharjee@imec.be}

	\author{Anupam Chattopadhyay}
	\affiliation{%
		\institution{School of Computer Science and Engineering,}
		\country{Nanyang Technological University, Singapore}
	}

	\author{Srijit Dutta}
	\affiliation{%
		\institution{Samsung Electronics}
		\city{Samsung Digital City}
		\country{South Korea}
	}
	
	\author{Ronny Ronen, Shahar Kvatinsky}
	\affiliation{%
		\institution{Andrew and Erna Viterbi Faculty of Electrical Engineering}
		\city{Technion-Israel Institute of Technology, Haifa}
		\country{Israel}}

	\renewcommand{\shortauthors}{Bhattacharjee et al.}
	
\begin{abstract}
Data-intensive applications are poised to benefit directly from processing-in-memory platforms, such as memristive Memory Processing Units, which allow leveraging data locality and performing stateful logic operations. Developing design automation flows for such platforms is a challenging and highly relevant research problem. In this work, we investigate the problem of minimizing delay under arbitrary area constraint for MAGIC-based in-memory computing platforms. We propose an end-to-end area constrained technology mapping framework, CONTRA. CONTRA uses Look-Up Table~(LUT) based mapping of the input function on the crossbar array to maximize parallel operations and uses a novel search technique to move data optimally inside the array. CONTRA supports benchmarks in a variety of formats, 
along with crossbar dimensions as input to generate MAGIC instructions. CONTRA scales for large benchmarks, as demonstrated by our experiments. CONTRA allows mapping benchmarks to smaller crossbar dimensions than achieved by any other technique before, while allowing a wide variety of area-delay trade-offs.  
CONTRA improves the composite metric of area-delay product by $2.1\times$ to $13.1\times$ compared to seven existing technology mapping approaches. 
\end{abstract}

\ccsdesc[500]{Hardware~Memory and dense storage}
\ccsdesc[500]{Software and its engineering~Source code generation}
	\keywords{In-memory computing, RRAM, Technology mapping, Design automation flow, MAGIC operations}
	
	\maketitle


\section{Introduction} 
The separation between the processing units and memory unit requires data transfer over energy-hungry buses. This data transfer bottleneck is popularly known as {\em the memory wall}. The overhead in terms of energy and delay associated with this transfer of data is considerably higher than the cost of the computation itself~\cite{pedram2016dark}. Extensive research has been conducted to overcome the memory wall, ranging from the classic memory hierarchy to the close integration of processing units within the memory~\cite{aga2017compute,seshadri2017ambit}. However, these methods still require transfer of data between the processing blocks and the memory, thus falling into the category of von Neumann architectures. 

Processing data within the memory has emerged as a promising alternative to the von Neumann architecture. This is generally referred to as {\em Logic-in-Memory}~(LiM). The primary approach to perform LiM is to store input variables or/and logic output in a memory cell. This is enabled when the physical capabilities of the memory can be used for data storage~(as memory) and computation~(as logic).  Various memory technologies, including Resistive RAM~(RRAM), Phase Change Memory~(PCM), Spin-transfer torque magnetic random-access memory~(STT-MRAM) and others have been used to realize LiM computation~\cite{lehtonen2009stateful,agrawal2018x, eike_logic,Gaillardon, kingra2020slim,kvatinsky2014magic,hamdioui2015memristor}. 

RRAM is one of the contending technologies for logic-in-memory computation. RRAMs permit {\em stateful logic}, where the logical states are represented as resistive state of the devices and at the same time, are capable of computation. Multiple functionally complete logic families have been successfully demonstrated using RRAM devices~\cite{reuben2017memristive}. In the following, three prominent logic families are presented.

\begin{figure}[t]
	\centering
	\includegraphics[width=\columnwidth]{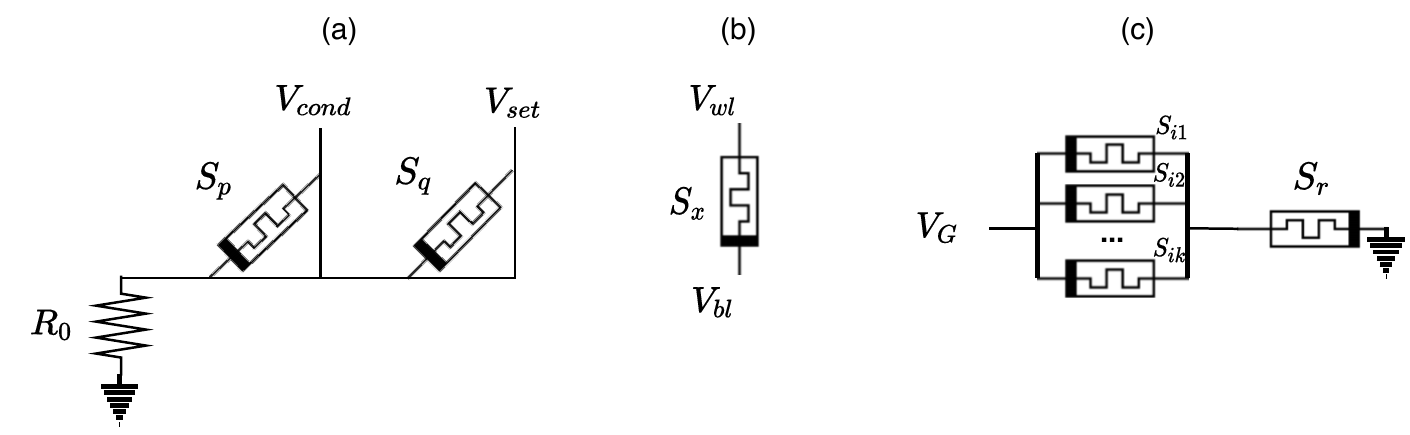}
	\caption{Logic primitives realized using memristors. (a) Material Implication (b) Majority logic (c) Memristor Aided Logic (MAGIC). }
	\label{fig:primitive}
\end{figure}

\noindent{\em Material Implication Logic}~\cite{lehtonen2009stateful}: Consider two RRAM devices~$p$ and $q$ with internal states $S_p$ and $S_q$ respectively, as shown in Fig.~\ref{fig:primitive}a. By applying voltages to the terminal, material implication can be computed, with the next state~(NS) of device~$p$ set to the result of computation. 
\begin{equation}NS_{p} = S_p \rightarrow S_q\end{equation}
\noindent{\em Majority Logic}~\cite{Gaillardon}: In this approach as shown in Fig.~\ref{fig:primitive}b, the wordline voltage~($V_{wl}$) and
bitline voltages~($V_{bl}$) act as logic inputs, while the internal resistive state $S_x$ of the device~$x$ acts a
third input. The next state of device $x$ in this case is a function of three inputs as shown below in the following equation.
\begin{equation}NS_{x} = M_3(S_x, V_{wl},\overline{V_{bl}})\end{equation}
\noindent{\em Memristor-Aided loGIC~(MAGIC)}~\cite{kvatinsky2014magic}. MAGIC allows in-memory compute operation by using the internal resistive state of single or multiple RRAM devices as input. The exact number of inputs~($k$) depends on the specific device used for computation. The result of computation is written to a new device~($r$), as shown in Fig.~\ref{fig:primitive}c. The internal resistive state of the input devices remain unchanged. Using MAGIC operations, multi-input NOR and NOT can be realized. 
\begin{align}
    NS_r &= NOR(S_{i1},S_{i2},\ldots,S_{ik}) \\ 
NS_r &= NOT(S_i) 
\end{align}

General purpose architectures have been proposed based on these primitives.  A bit-serial Programmable Logic in Memory~(PLiM) architecture was proposed by Gaillardon et al.~\cite{Gaillardon} that uses majority as the logic primitive. PLiM relied on using the same crossbar for storage of instructions as well for computation. RRAM-based Very long instruction word~(VLIW) Architecture for in-Memory comPuting (ReVAMP) was proposed by Bhattacharjee et al.~\cite{bhattacharjee2017revamp}, that used  Instruction Memory for the instruction storage and a separate RRAM crossbar as data storage and computation memory. 
Haj Ali et al. proposed memristive Memory Processing Unit~(mMPU)~\cite{haj2018not}. The mMPU consists of memristive memory arrays, along with Complementary Metal Oxide Semiconductor~(CMOS) periphery and control circuits to allow support for computations as well as conventional data read and write operations. To perform a computation within the mMPU, a compute command is sent to the mMPU controller. The controller generates the corresponding control signals and applies the signals to the crossbar array to perform the actual MAGIC operations. The mMPU allows MAGIC NOR and NOT gates to be executed within any part of the crossbar array, which allows storage of data as well as computation to happen in the same array. Compared to the architectures based on Material Implication, and Majority logic, MAGIC provides an inherent advantage. For MAGIC, control signals are not dependent on the output of a compute operation, . 

Wider acceptance of these architectures and technologies critically rely on efficient design automation flows, including logic synthesis and technology mapping. In this paper, we focus on the \textit{technology mapping challenge for architectures supporting MAGIC operations}. Intuitively, a Boolean function (represented using logic level intermediate form) is processed by the technology mapping flow to generate a sequence of MAGIC operations which are executed on the limited area available on a crossbar.  The number of devices available for computation using MAGIC operations on the mMPU is limited~\cite{lee20171,xue20130}, which makes the problem of technology mapping even more challenging. This particular variant is known as {\em area-constrained technology mapping problem}~(ACTMaP) for mMPU. Multiple technology mapping solutions for mMPU have been proposed in the literature~\cite{talati2016logic,thangkhiew2018scalable,hursimple,tenace2019said, yadav2019look,ben2019simpler}. Almost all of these works focus delay reduction, only one ~\cite{ben2019simpler} accepts a limited form of area constraints (limited row-size only) and considers device reuse to improve area efficiency.


In this paper, we propose CONTRA\footnote{Source code available: \url{https://github.com/debjyoti0891/arche}} -- the first scalable area-constrained technology mapping flow for the LiM computing using MAGIC operations. CONTRA not only allows specifying overall area constraint (in terms of number of devices) but also the exact crossbar dimensions. This enables CONTRA to map the same function into say a  $64\times64$ or $128\times128$ crossbar with different delays, whereas the existing methods cannot offer this flexibility. Specifically, our paper makes the following contributions:
\begin{itemize}
\item We propose a scalable 2-dimensional area-constrained technology mapping flow for the LiM computing using MAGIC operations. 
\item We present novel algorithms, using NOR-of-NORs representations (NoN) to place the LUTs on the crossbar to maximize parallelism, while maintaining the area constraints. We use an optimal A* search technique for moving inputs to the required position in the crossbar and propose an input alignment optimization to reduce the number of copy operations. 
\item We extensively evaluate our technique using various benchmarks. The overall flow achieves improvement in area-delay product from $2.1\times$ to $13.1\times$ in terms of geometric mean compared to seven existing technology mapping approaches for MAGIC. Our method can map arbitrary Boolean function using MAGIC operations to a smaller crossbar dimensions than achieved by any other technique before.
\end{itemize}
CONTRA takes an input benchmark, processes it using the novel technology mapping flow to generates MAGIC instructions. We developed an in-house simulator for MAGIC to execute the instructions and formally verify the functional equivalence of the generated instructions and the input benchmark. 

\section{Background and Related Works}\label{sec:background}
\subsection{ MAGIC operations}
\begin{figure}[t]
    \begin{subfigure}[b]{\columnwidth}
    \centering
		\includegraphics[width=0.6\columnwidth]{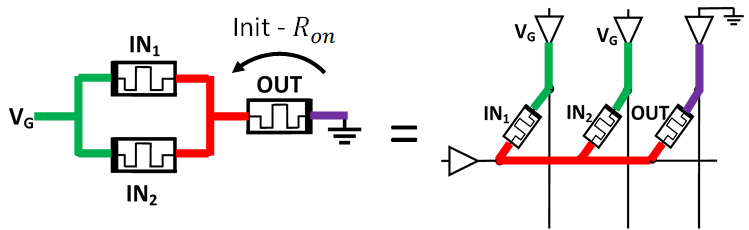}
		\caption{MAGIC  operations using memristors, which can be performed in a crossbar configuration.}
		\label{fig:magic_op}
	\end{subfigure}
	
	\centering
	\begin{subfigure}[b]{0.45\columnwidth}
		\includegraphics[width=0.6\columnwidth]{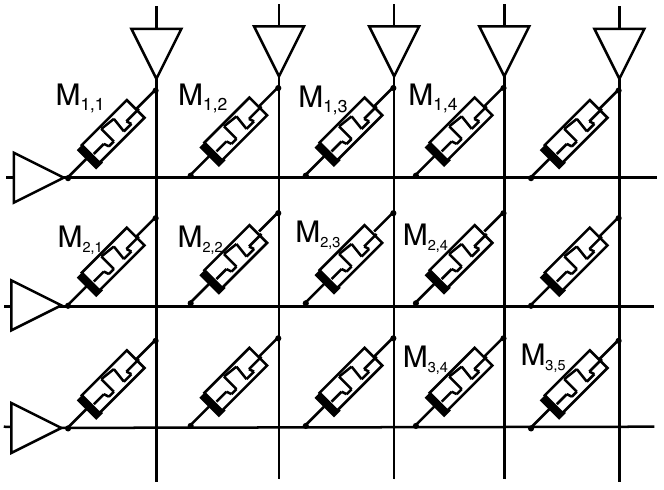}
		\caption{Memristors arranged in a crossbar configuration.}
		\label{fig:mem_grid}
	\end{subfigure}
	\centering
	\begin{subfigure}[b]{0.45\columnwidth}
		\centering
		\includegraphics[width=\columnwidth]{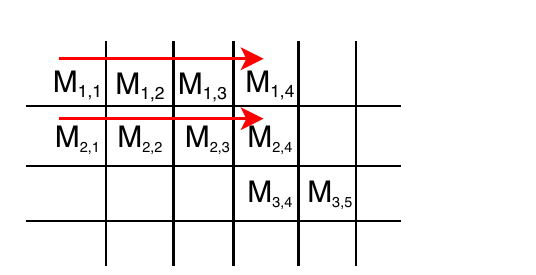}
		\caption{Horizontal NOR}
		\label{fig:action1}
	\end{subfigure}
	
	\begin{subfigure}[b]{0.45\columnwidth}
		\centering
		\includegraphics[width=\columnwidth]{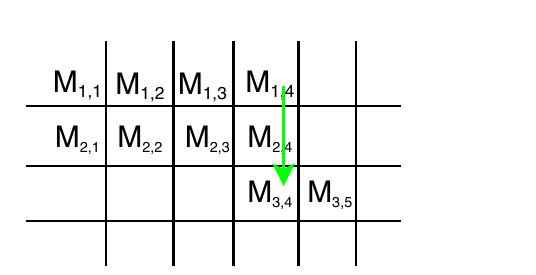}
		\caption{Vertical NOR}
		\label{fig:action2}
	\end{subfigure}
	\hfill
	\begin{subfigure}[b]{0.45\columnwidth}
		\centering
		\includegraphics[width=\columnwidth]{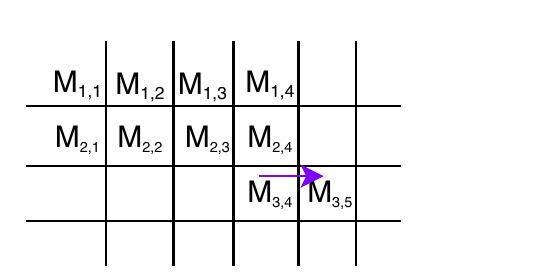}
		\caption{NOT}
		\label{fig:action3}
	\end{subfigure}
	\caption{Basic MAGIC operations on a crossbar array.}
	\label{fig:NoN}
\end{figure}

We present the basics of computing using MAGIC operations to begin with. 
As shown in Fig.~\ref{fig:magic_op}, a 2-input MAGIC NOR gate consists of 2-input memristors~($IN_1$ and $IN_2$) and one output memristor~($OUT$). The memristive state of the output memristor changes in accordance with the resistive states of the input memristors. Low resistive state is interpreted as logical `1' while high resistive state is interpreted as logical `0'. The NOR gate operation is realized by applying $V_{G}$ to the input memristors while the output memristor is grounded. Note that the output memristor has to be initialized to low resistive state before the NOR operation is carried out. After applying the voltage, the resistance of the output memristor is set based on the ratio between the resistances of the input and the output memristors and results in a NOR operation. The MAGIC NOR operation can be performed with the devices arranged in a crossbar configuration, as shown in the right hand side of Fig.~\ref{fig:magic_op}. By extending this approach, it is feasible to perform logical $n$-input NOR and NOT operations. 

 
Multiple MAGIC operations can be performed in parallel.  The parallel execution of multiple NOR gates is achieved whenever inputs and outputs of the $n$-input NOR gates are aligned in the same rows or columns in a crossbar, as shown in Fig.~\ref{fig:mem_grid}. For example, Fig.~\ref{fig:action1}, two 3-input NOR operations are performed in parallel. 
\begin{align} 
M_{1,4} = NOR(M_{1,1}, M_{1,2},M_{1,3})  M_{2,4} = NOR(M_{2,1}, M_{2,2}, M_{2,3}) \nonumber  
\end{align}
Also, vertical operations are allowed as shown in Fig.~\ref{fig:action2}.
$$M_{3,4} = NOR(M_{1,4}, M_{2,4})$$
A single-input NOR operation  is  a  NOT  gate, as shown in Fig.~\ref{fig:action3}.
$$M_{3,5} = NOT(M_{3,4})$$ 
Thus, both $n$-input NOR and NOT gates can be executed by MAGIC operations. It is also possible to reset the devices in parallel in the crossbar to `1', either row-wise or column-wise. 

\subsection{Logic Synthesis and Technology Mapping}
\begin{figure}[t]
	\centering
	\includegraphics[width=0.9\columnwidth]{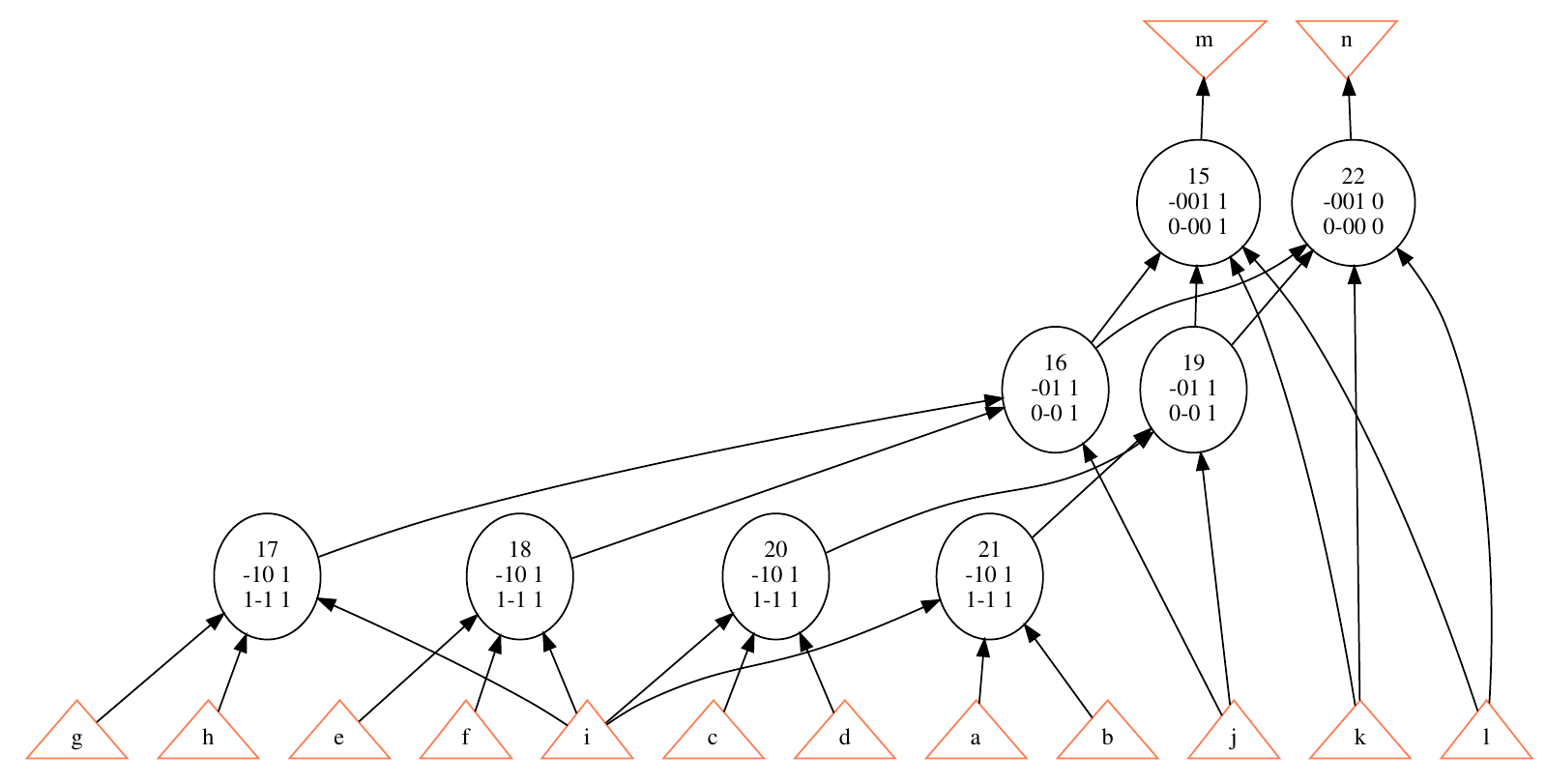}
	\caption{\texttt{cm151a}~benchmark partitioned into LUTs with $k=4$. Each triangular node represents a primary input, while the inverted triangle represent primary outputs. Each round node represents a LUT. LUT id and their functionality in SoP is shown inside the node.}
	\label{fig:lutgraph}
\end{figure}

For logic synthesis and technology mapping approaches, a classification of different Intermediate Representations (IRs) has been proposed in~\cite{soeken2016unlocking}. First, there are \textit{Functional} approaches, where the IR is used to explicitly express the logic function. Examples for IRs are Boolean truth tables, Look-Up Tables (LUTs) or Binary Decision Diagrams (BDDs). Second, there are \textit{Structural} approaches, where the IR is used to represent the structure of the circuit, e.g., using And-Inverter Graphs (AIGs). For technology mapping on memristive crossbar, both types of approaches have been adopted, as it fits more closely the device-level operations. Among the design automation flows developed for memristive technologies, Majority-Inverter Graphs (MIGs) are predominantly used due to their native mapping on to devices supporting Majority Boolean functions~\cite{bhattacharjee2018technology,shirin2018TCAD}. MAGIC devices realize multi-input NOR operations, which do not allow a direct mapping from MIGs. Hence, in this work, we use LUT graph and NOR-of-NOR representations for solving ACTMaP for mMPU. The rationale for using LUT graph is that it allows mapping to all forms of Boolean functions~\cite{tenace2019said}. \\
\noindent \textbf{LUT graph:} Any arbitrary Boolean function can be represented as a directed acyclic graph~(DAG) $G = \langle V,E\rangle$, with each vertex having at most $k$-predecessors~\cite{abc}. Each vertex $v$, $v \in V$, with $k$-predecessors represents a $k$-input Boolean function or simply a $k$-input LUT. Each edge, $u \rightarrow v$ represents a data dependency from the output of node $u$ to an input of node $v$. 
\begin{example}
	Fig.~\ref{fig:lutgraph} shows the cm151a benchmark from LGSynth91 as a DAG with $k=4$. The benchmark has 12 primary inputs $a-l$ and two primary outputs $m$ and $n$.  LUT $16$ has a dependency on LUTs $17$ and $18$ and on primary input~$j$. We use this benchmark as a running example to explain the proposed method.
\end{example}
\begin{figure*}[t]
	\centering
	\includegraphics[width=0.65\textwidth]{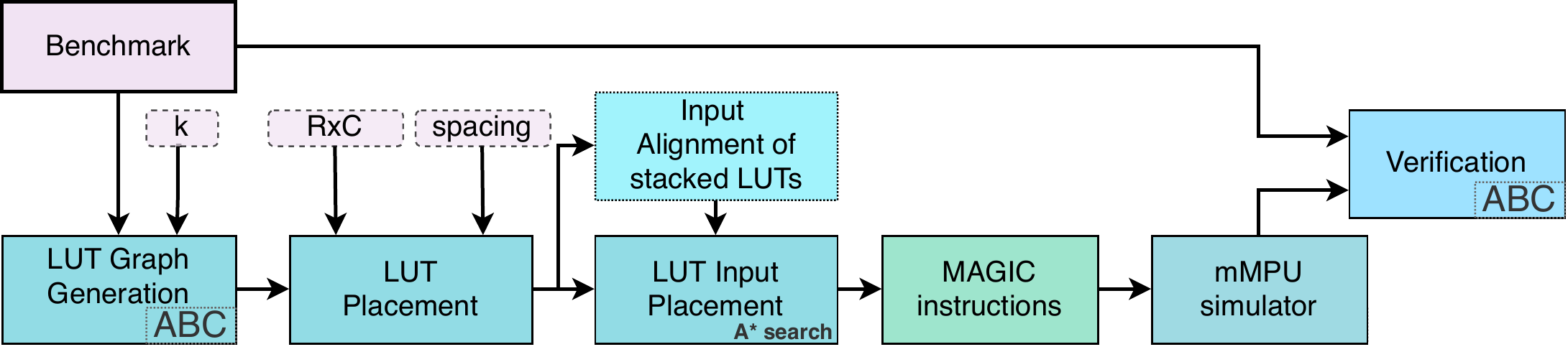}\qquad
	\caption{CONTRA: area-\underline{Con}strained \underline{T}echnology mapping f\underline{RA}mework for Memristive Memory Processing Unit}
	\label{fig:sacflow}
	\vspace{-0.5cm}
\end{figure*}

\noindent \textbf{NOR-of-NOR representation}: A Boolean function $\mathcal{F} : \mathcal{B}^n \rightarrow \mathcal{B}$, expressed in sum-of-products (SoP) form can be converted to the NOR-of-NORs~(NoN) representation by the following simple transformations. 
\begin{enumerate}
	\item Replace $\vee$ and $\wedge$ operations with $\overline{\vee}$
	\item Flip the polarity of each primary input
	\item Negate the result
\end{enumerate}
For example, we can express $F$ in NoN representation as follows.  
\begin{align}
F &= (a \wedge \overline{b}) \vee  (\overline{a} \wedge b \wedge c)
&= \overline{(\overline{a}\ \overline{\vee}\ b)\ \overline{\vee}\  (a\ \overline{\vee}\ \overline{b}\ \overline{\vee}\ \overline{c})} \label{eq:F}
\end{align}
Alternatively, we can express this NoN as:-
\begin{center}
	\begin{tabular}{lccc}\hline 
		Variables & $a$ & $b$ & $c$ \\\hline
		1st product term: & 0 & 1 & - \\
		2nd product term: & 1 & 0 & 0 \\\hline
	\end{tabular}
\end{center}


\subsection{Related Works}
Multiple works address the issue of design automation for computation with bound on the number of memristive devices. Lehtonen et al. presented a methodology for computing arbitrary Boolean functions using devices that realize material implication~\cite{lehtonen2009stateful}. For any Boolean function with $n$~inputs and $m$~outputs, $m + 2$ working memristors are required for computing the function. For $n$-input Boolean function with a single output, three working memristors are sufficient for computation. This bound was further reduced to two working memristors by Poikonen et al.~\cite{lehtonen2010two}. Optimal and heuristic solutions for ACTMaP for devices realizing majority with single input inverted have been proposed in~\cite{aomap}. Crossbar-constrained ACTMaP solution has been proposed for devices realizing majority with single input inverted in~\cite{bhattacharjee2018crossbar}. 

As mentioned before, several technology mapping methods for mMPU have been proposed in literature~\cite{talati2016logic,thangkhiew2018scalable,hursimple,tenace2019said, yadav2019look,ben2019simpler}. 
These methods primarily work towards reducing latency for mapping an arbitrary function and output the dimensions of the crossbar required to map the function. While trying to maximize parallelism, these methods often map to highly skewed crossbar dimensions (where number of rows is much higher than number of columns or vice versa).  Furthermore, this methods are highly area inefficient since they do not reuse devices, leading to very low device utilization. 
To our knowledge, SIMPLER~\cite{ben2019simpler} is currently the only method for mMPU that is optimized for area. 
SIMPLER relies on mapping functions to a single row, with the objective of achieving high throughput by simultaneously executing multiple data streams in different rows. As SIMPLER allows device reuse, it has high area utilization. However, the utility of this method is limited as all the used devices must still be allocated in a single memory row and it cannot use 2-dimensional crossbar for mapping in order to fit a function into a small crossbar. We address the challenge of \mbox{2-dimensional} constrained mapping.

%
%

\section{Area-constrained Technology Mapping Flow}\label{sec:method}
In this section, we describe CONTRA, a 2-dimensional area-\underline{Con}strained \underline{T}echnology mapping f\underline{RA}mework for memristive memory processing unit, which is shown in Fig.~\ref{fig:sacflow}.


\subsection{LUT Placement on Crossbar}
The goal of this phase is to map the individual nodes~(LUTs) of the input DAG on the crossbar, so as to minimize the delay of computing. LUTs in the same topological level of the DAG do not have any dependencies between themselves and therefore, could be scheduled in parallel. In order to permit computation of multiple LUTs in parallel, we utilize the NOR-of-NOR representation of the LUT function.

Since the NoN representation consists of only NOR and NOT operations, it can be computed by MAGIC operations directly in $3$ cycles, ignoring the initialization cycle(s). All the variables in appropriate polarity~(inverted or regular) in a product term are aligned in rows. For the variables which are not present in a product term, the corresponding memristor is set to `1', which is the state of the memristor after reset. This is followed by computing NOR of all the product terms horizontally in a single cycle. In the next cycle, a vertical NOR of the above results produces the negated output. In the last cycle, we negate this result to get output of the computed function.
\begin{figure}[ht]
	\includegraphics[width=\columnwidth]{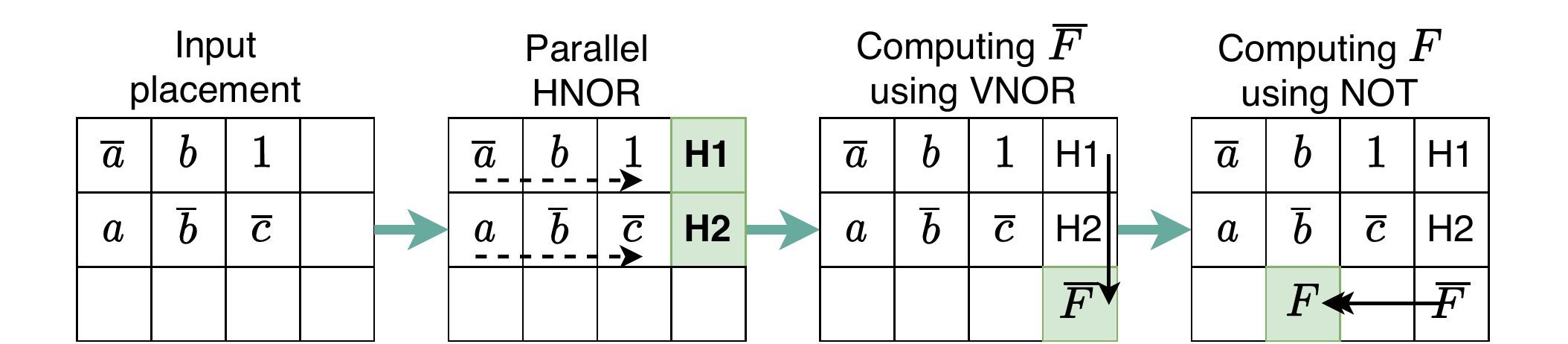}
	\caption{Computation of $F$ with 3 inputs and 2 product using MAGIC operations on a $3\times4$ crossbar.}
	\label{fig:noncomp}
 		\vspace{-0.4cm}
\end{figure}
\begin{example}
	The computation of $F$ in equation~\eqref{eq:F} using MAGIC operations is shown Fig.~\ref{fig:noncomp}. Row 1 and row 2 have the inputs for the 1st and 2nd product terms respectively. These inputs are NORed in parallel to compute the product terms with the outputs written to $M_{1,4}$~(H1) and $M_{2,4}$~(H2). The product terms are vertically NORed to compute~$\overline{F}$ in $M_{3,4}$. In the final step, $\overline{F}$ is inverted using a NOT operation to compute~$F$ (in $M_{3,2}$).  
\end{example}

The LUTs are topologically ordered and grouped by the number of inputs. The LUTs are  placed one below another with inputs aligned till we are limited by the height of the crossbar. Consider $n$-LUTs each with $k$-inputs. Once the LUTs are aligned one below another, we can compute the horizontal NOR of all LUTs in one cycle. This is because the inputs and outputs of all the LUTs are aligned and the voltage of the columns applies to all LUTs. In the next $n$-cycles, we can perform the vertical NOR operations to compute the inverted output of the $n$ stacked LUTs. Thus, $(n+1)$ cycles are required to compute the $n$ stacked LUTs. Let us consider that each $k$-input LUT $L_i$ has $p_i$ product terms, $1 \le i \le n$. Then, the area $L^n_{area}$ required to compute the $n$ LUTs in parallel is :-
\begin{equation}
L^n_{area} = \sum_{i=1}^{n} (p_i+1)\times(k+1)
\end{equation}
The LUT placement strategy is from top to bottom and from left to right. The $spacing$ parameter is used to specify the number of rows that are left empty between two LUTs stacked vertically. If we do not have enough free devices to place a new LUT, the  crossbar is scanned row-wise and column-wise to check in which rows or columns, the intermediate results are present. These are considered blocked and the rest of the crossbar is reset either row-wise or column-wise, which results in lesser number of devices being blocked. . The process is repeated till all the LUTs are placed. The overall flow is presented in Algorithm~\ref{algo:lutplacement}.

\begin{figure}[t]
	\centering
	\includegraphics[width=\columnwidth]{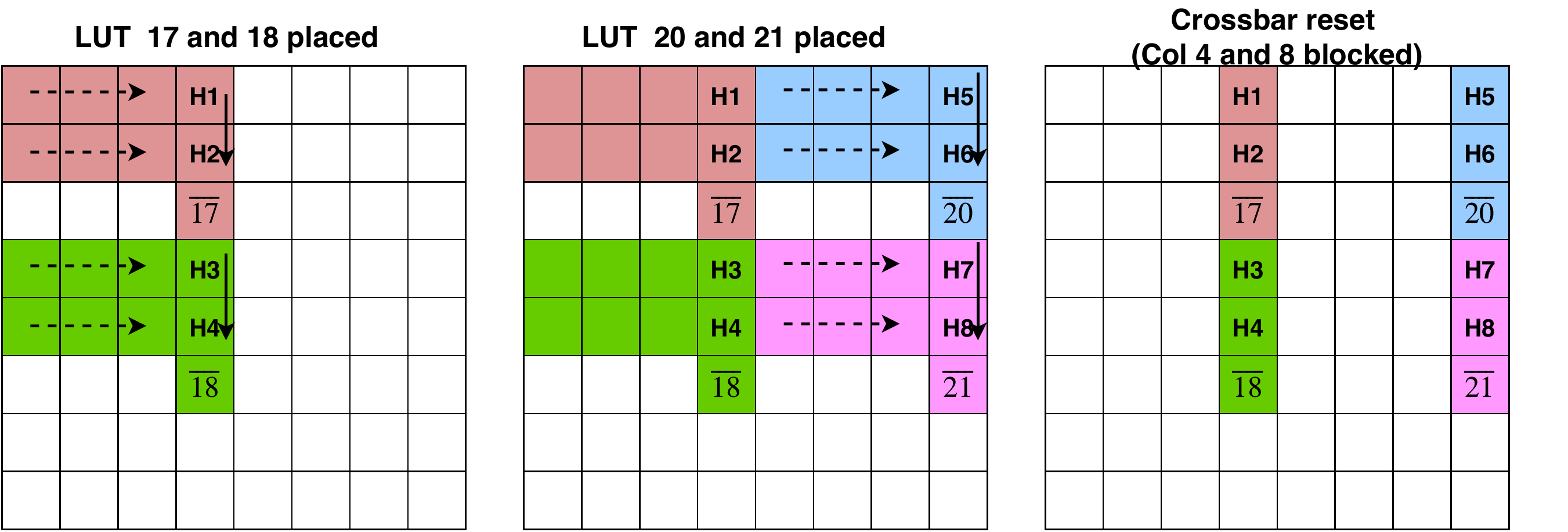}\vspace{0.1cm}
	\caption{LUT Placement phase on a $8\times8$ crossbar for the cm151a benchmark.}
	\label{fig:cm151crossbar}
	\vspace{-0.2cm}
\end{figure}

\begin{example}
For cm151a, we stack the LUTs $17$ and $18$ in the crossbar, as shown in Fig.~\ref{fig:cm151crossbar}. Since enough space in not available vertically, we stack LUTs $20$ and $21$ on the right. We reset the crossbar, without resetting column $4$ and $8$, as these columns contain the intermediate results. We continue placing the other LUTs in similar manner. 
\end{example}
\text{
	\begin{tabular}{rl}
		1 & [LUT 17 (1,1)$\rightarrow$(3,4) ][ LUT 18 (4,1)$\rightarrow$(6,4) ] \\
		2 & [LUT 20 (1,5)$\rightarrow$(3,8) ][ LUT 21 (4,5)$\rightarrow$(6,8) ] \\
		3 & [Reset columns except \{4,8\}] \\
		4 & $\dots$ \\
	\end{tabular}
}

Note that we are effectively computing the inverted output of each LUT. Therefore, for the output LUTs, an additional NOT operation is required, as specified in lines 17-19 of Algorithm~\ref{algo:lutplacement}.

\begin{algorithm}[h]
	\footnotesize 
	\SetAlgoLined
	\SetKwInOut{Input}{Input}\SetKwInOut{Output}{Output}
	\Input{$G$, $R$, $C$, $spacing$}
	\Output{Mapping of $G$ to crossbar $R\times C$.}
	\BlankLine
	
	\Do{There is a LUT not yet placed.}{
		$L_{set}$ = Pick LUTs in a topological level with equal number of inputs.  
				
		\If{limited by space vertically}{
			Start placing from next available column \; 
		}
		\If{limited by both vertical and horizontal space}{Reset the cells keeping the intermediate outputs intact.}
		Place $L_{set}$ stacked together vertically with $spacing$ rows empty between subsequent LUTs.\\
		Schedule all the LUTs in $L_{set}$ in the same time slot of the  schedule.
	}

	\For{Each set of LUTs stacked together}{
	Place the inputs for these LUTs, using A* search and vertical copies\;
	Compute intermediate results in parallel using Horizontal NORs.\;
	Compute inverted output of LUTs in sequence using Vertical NORs.\;
	}
	
	\For{Each inverted output of $G$}
	{ Invert using NOT operation to compute outputs of $G$.}
	
	\caption{Area-constrained technology mapping.}
	\label{algo:lutplacement}
\end{algorithm}

\subsection{LUT Input Placement Technique}
For some of the LUTs,, we require the intermediate outputs from previous computations as inputs. We use $A^*$ search to get the shortest path to copy an intermediate value from source~($R_S,C_S$) to destination~($R_D,C_D$) with a minimum number of NOT operations. The cost of a location $cost(r,c)$ is $f(r,c) + g(r,c)$. $f(r,c)$ is equal to the number of copy operations used to reach from ($R_S,C_S$) till $(r,c)$. 
\begin{equation}
g(r,c)=
\begin{cases}
0, \text{if $(r,c)$ is the destination}\\
1, \text{if } r == R_D \text{ or } c == R_C\\
2, \text{otherwise}
\end{cases}
\end{equation}

\begin{figure}
    \centering
    \includegraphics[width=\columnwidth]{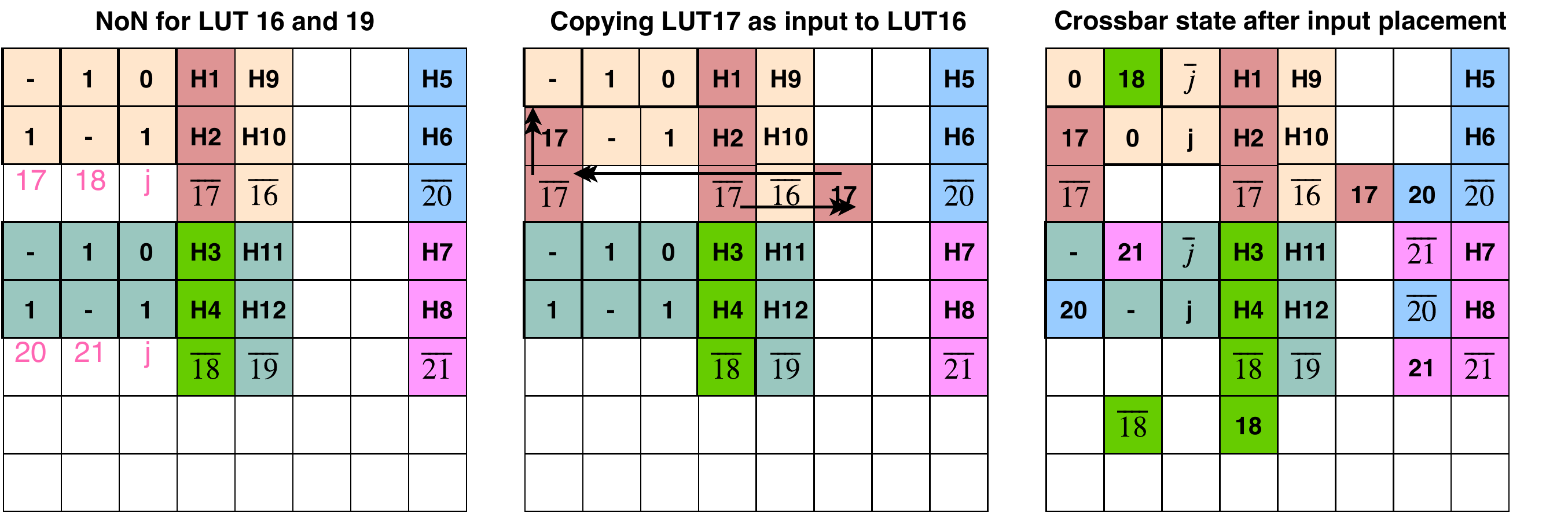}
    \caption{Placement of the inputs for LUTs 16 and 17 and the corresponding literals for NOR-of-NOR computation.}
    \label{fig:aligneg}
\end{figure}

All empty cells in the row and column of the current location are considered its neighbours. The search starts at the source, updates the cost of the neighbouring location and picks the location with the least $cost$. The process is repeated till the goal state is reached. If the path length is odd, the polarity of the input is reversed while for an even length path, the polarity is preserved. This is due to an odd or even number of NOT operations respectively. If the inputs of a NoN has only positive or negative terms, but not both, we need to choose the copy path to be even or odd accordingly. If the inputs are of mixed polarity, we can choose the path with shorter length, the polarity does not matter. Thereafter, the input variable is vertically copied to different rows as required for the other product terms in the LUT, according to the NoN representation.	\begin{example}
	LUT 16 uses the output of LUT 17 as input, with the NoN representation shown in Fig.~\ref{fig:aligneg}. We copy the value from $M_{3,4}$ to $M_{3,1}$ using a sequence of NOT operations, obtained using $A^*$ search.
	\begin{center}\footnotesize NOT($M_{3,4}\rightarrow M_{3,6}$), NOT($M_{3,6}\rightarrow M_{3,1}$), NOT($M_{3,1}\rightarrow M_{2,1}$)\end{center}
	The state of the crossbar after placing all the inputs~(LUTs 17, 18, 20 and 21, primary inputs i and j) for LUT~$16$ and $19$ is shown in the last sub-figure of  Fig.~\ref{fig:aligneg}.
\end{example}

%
%
%
%

\subsection{Input Alignment for multiple LUTs}
Multiple LUTs scheduled together for execution, often share common inputs. If the common inputs are assigned to the same column, then only a single $A^*$ search would be required to bring the input to the column, and followed by vertically copying to the appropriate rows. This would lead to reduction in delay as well as reduction in the number of devices involved in copying. The goal is to have an assignment of the inputs of the individual LUTs to columns such that it maximizes the number of aligned inputs in a set of stacked LUTs. 

We encode the constraints of this problem  to optimally solve the problem using an Satisfiability Modulo Theories~(SMT) solver. \\
$\rhd$ \texttt{Maximize} $\sum_{c=1}^k \sum_{li=1}^n \sum_{lj=1}^n align_{c,li,lj}$ \\
$\rhd$ $A_{c,l} = v | \exists v \in $ input of LUT~$l$. $1 \le c \le k$ and $1\le l \le n$.\\ 
$\rhd$ $align_{c,li,lj} = 1$ if $A_{c,li} = A_{c,lj}$.  $1 \le c \le k$, $1\le li \le n$  and $1\le lj \le n$. \\
The assignment to variable $A_{c,l} = v$ indicates a variable~$v$ is assigned to column~$c$ of LUT~$l$. For $n$ LUTs each with $k$ inputs, a brute force approach would have time complexity of $(k!)^{n-1}$. As the SMT solver takes a long time to solve and have to be executed multiple times in mapping a benchmark, we propose a  greedy algorithm for faster mapping. 

Consider $k$-input LUTs and $n$ of these LUTs stacked together. This can be represented as a matrix with dimensions $n \times k$, where each row represents the inputs variables of the LUT. As the inputs of an LUT are unique, each variable occurs at most once in each row of the matrix. The detailed alignment approach is shown in Algorithm~\ref{algo:input_align}. We explain the algorithm with a representative example.

\begin{algorithm}[t]
	\footnotesize
\SetKwInOut{Input}{Input}\SetKwInOut{Output}{Output}
\Input{$M$}
\Output{$M_{align}$}
\BlankLine
\SetAlgoLined
L = Ordered List of variables in the matrix in descending order by count.\\
$M_{align}$ = initialize $n \times k$ matrix with $\phi$; 

\For{variable $v$ in $L$}
{ 
  $R_v$ = $\{r$ if $v \in$ row $r$ of $M\}$;\\
  $target_c$ = None;\\
  \For{col $c$ in matrix $M$}
  {
     \If{ $M_{align}[r][c] == \phi | \forall r \in R_v$}
     {
        $target_c$ = $c$;\\
        break;
     }
  }
  \uIf{$target_c$ == None}
  {
    Place $v$ in any free column in each row  $\in R_v$; 
  }
  \Else{
    Place $v$ in column~$target_c$ in each row  $\in R_v$;  
  } 
}

   return $M_{align}$;

 \caption{Input Alignment}
 \label{algo:input_align}
\end{algorithm}

\begin{figure}[ht]
    \centering
    \includegraphics[width=0.7\columnwidth]{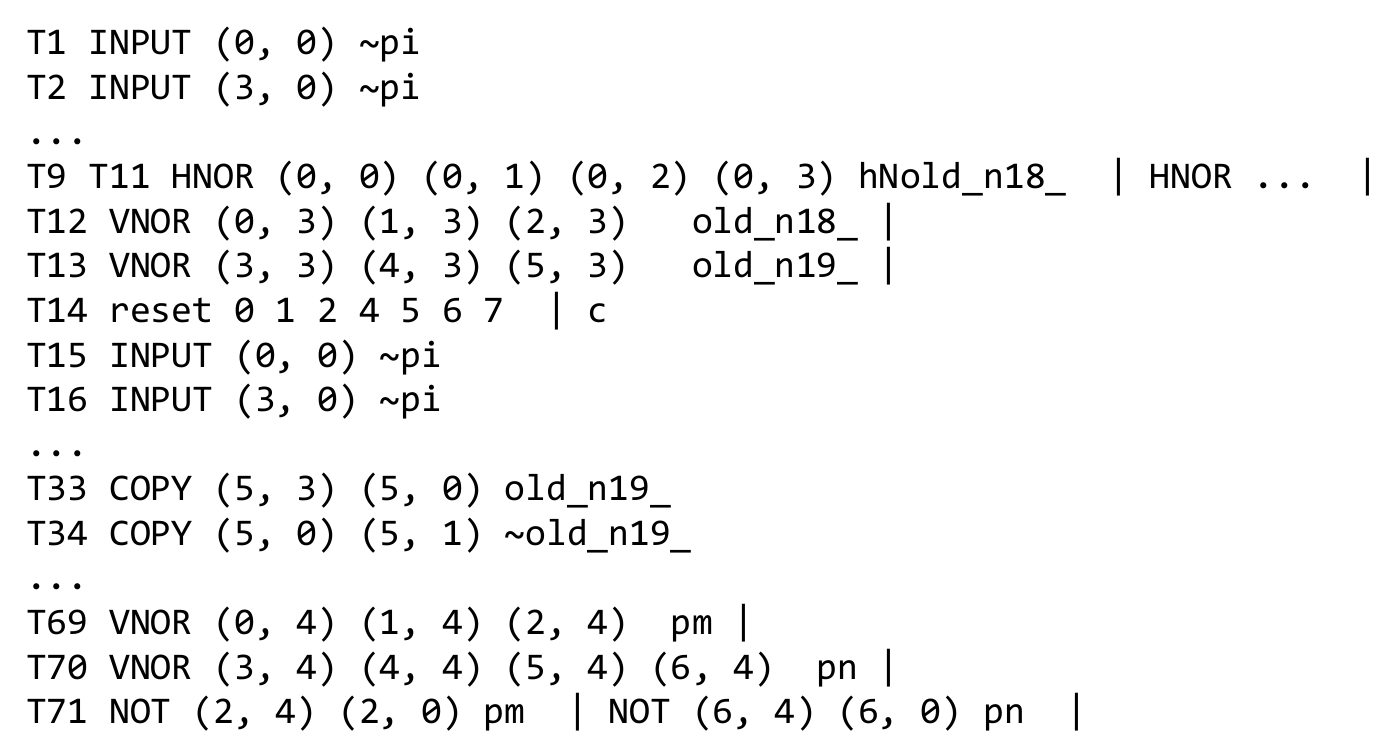}
    \caption{Snippet of MAGIC instructions generated by CONTRA on mapping \texttt{cm151a} benchmark on  $8\times8$ crossbar with $k=4$ and spacing=0.}
    \label{fig:contra_out}
    \vspace{-0.4cm}
\end{figure}

\begin{example}
Consider the three 4-input LUTs with their input variables arranged as an unaligned matrix, as shown below. The variables are ordered in descending order by frequency.
L = \{a:3, b:2, c:2, d:1, e:1, h:1, g:1, x:1\}.
We start the alignment by placing `a' in the first column. In the next step, we place `b'. As row 1 and 2 of column 1 are already occupied by `a', we place `b' in column 2. Similarly, we continue the process until all the variables are placed.

{\setlength{\tabcolsep}{2pt}
	\footnotesize
\begin{tabular}{cccc c cccc c cccc c cccc}
\multicolumn{4}{c}{\textbf{Unaligned}} & \multicolumn{4}{c}{\textbf{Step 1}} & &\multicolumn{4}{c}{\textbf{Step 2}} & ... &\multicolumn{4}{c}{\textbf{Aligned}} \\
a & b & c & d & & a & $\phi$ &$\phi$ & $\phi$ & & a & b &$\phi$ & $\phi$ & & a & b & c & d\\
b & c & e & a & &a & $\phi$ &$\phi$ & $\phi$ & & a & b &$\phi$ & $\phi$ & & a & b & c & e\\
h & a & g & x & & a & $\phi$ &$\phi$ & $\phi$ & & a & $\phi$ &$\phi$&$\phi$ &     & a & h & g & x\\
\end{tabular}
}
\end{example}
\begin{example}
For the LUTs 16 and 19, the result of alignment is shown in first sub-figure of Fig.~\ref{fig:aligneg}, specified by variables in pink. The variables 17, 18 and j are assigned columns $1$, $2$ and $3$ for LUT 17 while the variables 20, 21 and j are assigned columns $1$, $2$ and $3$ for LUT 18, thereby aligning input variable $j$.
\end{example}
This completes the description of the technique for area-constrained mapping. The output of mapping \texttt{cm151a} benchmark to $8\times8$ crossbar with $k=4$ and $spacing=0$ and  is shown in Fig.~\ref{fig:contra_out}.  The benchmark was mapped in 71 cycles. Each line signifies one or more operations with the corresponding input and gate names (pi, old\_n\_18, etc.) that are executed in the same cycle.  In the next section, we present the results of benchmarking the proposed method.
\section{Experimental Results}\label{sec:results}
\noindent This section presents the experimental results of the CONTRA, the proposed area-\underline{Con}strained \underline{T}echnology mapping f\underline{RA}mework for 
for computing arbitrary functions using MAGIC operations.  We have implemented the proposed CONTRA framwork using Python. CONTRA supports a variety of input formats for the benchmarks, including {\em blif}, structural {\em verilog}, {\em aig}.  We have used ABC~\cite{abc} for all generating the LUT graph and the SOP representation of LUT functions, which we converted to NoN representation for mapping. For each benchmark, CONTRA generates cycle accurate MAGIC instructions. A representative output of mapping is shown in Fig.~\ref{fig:contra_out}. We developed an in-house mMPU simulator for executing MAGIC instructions. We used the simulator to generate execution traces which were converted into Verilog. The generated Verilog and the input benchmarks were formally checked for functional equivalence using the {\em cec} command of ABC.

\begin{table}[ht]
    \setlength{\tabcolsep}{2pt}
    \caption{Benchmarking results for the ISCAS85 benchmark for three crossbar sizes. We ran each benchmark with \mbox{$k=\{2,3,4\}$} and spacing set to $\{0,2,4,6\}$. For each benchmark, the best results were obtained for $k=4$ and spacing set to $6$.}
    \label{table:iscas_res}
    \centering
\setlength{\tabcolsep}{3pt}
 \scriptsize
\begin{tabular}{rcc|r|r|rrc}\bottomrule
&  & (R,C) & \textbf{(64,64)} & \textbf{(128,64)} & \textbf{(128,128)} \\ 
\textbf{Bench} & \textbf{PI} & \textbf{PO} & \textbf{Cycles} & \textbf{Cycles} & \textbf{Cycles} \\  \midrule
c432  & 36  & 7  & 797  & 774  & 770   & \\
c499  & 41  & 32  & 1391  & 1341  & 1343   & \\
c880  & 60  & 26  & 1314  & 1268  & 1263    & \\
c1355  & 41  & 32  & 1390  & 1341  & 1344   & \\
c1908  & 33  & 25  & 1511  & 1470  & 1469   & \\
c2670  & 233  & 140  & 2132  & 2066  & 2060   & \\
c3540  & 50  & 22  & 3751  & 3575  & 3575   & \\
c5315  & 178  & 123  & 5022  & 4827  & 4831    & \\
c6288  & 32  & 32  & 8176  & 7890  & 7881   & \\
c7552  & 207  & 108  & 7308  & 7039  & 7036    & \\  \toprule
\end{tabular}
\vspace{-0.6cm}
\end{table}
\begin{table}[ht]
\centering 
\caption{Benchmarking results for the EPFL MIG benchmarks for three crossbar sizes. We ran each benchmark with \mbox{$k=\{2,3,4\}$} and spacing set to $6$. For each benchmark, the best results were obtained for $k=4$.}
    \label{table:epfl_mig}
\setlength{\tabcolsep}{3pt}
 \scriptsize
\begin{tabular}{rcc|r|r|r}\bottomrule
&  & (R,C) & \textbf{(64,64)} & \textbf{(128,64)} &\textbf{(128,128)} \\ 
\textbf{Bench} & \textbf{PI} & \textbf{PO} & \textbf{Cycles} & \textbf{Cycles} & \textbf{Cycles} \\  \midrule
arbiter & 256 & 129 & 81941 & 81582 & 81434 \\
cavlc & 10 & 11 & 3808 & 3672 & 3686 \\
ctrl & 7 & 26 & 786 & 759 & 757 \\
dec & 8 & 256 & 1399 & 1253 & 1284 \\
i2c & 147 & 142 & 6698 & 6656 & 6692 \\
int2float & 11 & 7 & 1369 & 1340 & 1323 \\
priority & 128 & 8 & 5479 & 5398 & 5389 \\
router & 60 & 30 & 1150 & 1121 & 1153 \\
voter & 1001 & 1 & - & 68777 & 68758 \\ \toprule
\end{tabular}
 \vspace{-0.6cm}
\end{table}

\begin{table}[t]
\centering
\caption{Benchmarking results for the EPFL arithmetic benchmarks for $256\times256$  crossbar size.}
\label{table:epfl_arith}
\setlength{\tabcolsep}{3pt}
 \scriptsize
\begin{tabular}{rcc|r|ccrcc}\bottomrule
\textbf{Bench} & \textbf{PI} & \textbf{PO} & \textbf{LUTs} &  \textbf{k} & \textbf{Spacing} & \textbf{Cycles} \\ \midrule
adder & 256 & 129 & 339 & 4 & 6 & 4398 & \\
bar & 135 & 128 & 1408 & 4 & 6 & 12216 & \\
div & 128 & 128 & 57239 & 2 & 6 & 342330 & \\
hyp & 256 & 128 & 64228 & 4 & - & -\\
log2 & 32 & 32 & 10127 & 4 & 1 & 128647 & \\
max & 512 & 130 & 1057 & 4 & 6 & 9468 & \\
multiplier & 128 & 128 & 10183 & 3 & 0 & 90925 & \\
sin & 24 & 25 & 1915 & 4 & 6 & 21761 & \\
sqrt & 128 & 64 & 8399 & 4 & 6 & 101694 & \\
square & 64 & 128 & 6292 & 4 & 0 & 74614 & \\ \toprule
\end{tabular}
\vspace{-0.2cm}
\end{table}

We benchmark our tool using the ISCAS85 benchmarks~\cite{hansen1999unveiling}, which have been used extensively for evaluation of automation flows for MAGIC. The experiments were run on a shared cluster with 16  Intel(R) Xeon(R) CPU E5-2667 v4 @ 3.20GHz, with Red Hat Enterprise Linux 7. Table~\ref{table:iscas_res} shows the results of mapping the benchmarks for three crossbar dimensions.   We report the execution time in seconds for $128\times128$ for the ISCAS85 benchmarks. We report the results for the best delay~(in cycles) by varying $k$ from $2$ to $4$. As expected, the increase in crossbar dimensions results in lower delay of execution. We also report the results of mapping for the EPFL benchmarks\footnote{https://github.com/lsils/benchmarks}.  We report the results for EPFL MIG benchmarks in Table~\ref{table:epfl_mig} for three crossbar dimensions. For the larger EPFL arithmetic and random control benchmarks, we report the results for crossbar with $256\times256$ dimensions in Table~\ref{table:epfl_arith} and Table~\ref{table:epfl_random_control} respectively. 

We observe that for most of the results, the best delay was obtained for $k=4$. This is because setting a higher value of k, leads to fewer LUTs in the LUT graph. Since multiple LUTs can be scheduled in parallel (based on constraints mentioned in Algorithm~\ref{algo:lutplacement}), this leads to reduction in the number of cycles to compute the benchmark by exploiting higher degree of parallelism. For large benchmark such as {\em voter} in Table~\ref{table:epfl_mig} and very small crossbar dimension~($64,64$), the mapping flow fails. This happens because during the placement phase of the flow, multiple columns are blocked with intermediate results which does not leave enough number of free devices to map the rest of the LUTs.

\begin{table}[t]
\centering
\caption{Benchmarking results for the EPFL control benchmarks for $256\times256$ crossbar size, with spacing set to 6. We ran each benchmark with \mbox{$k=\{2,3,4\}$} and the best results were obtained for $k=4$.}
\label{table:epfl_random_control}
\setlength{\tabcolsep}{3pt}
 \scriptsize
\begin{tabular}{rcc|c|c}\bottomrule
\textbf{Benchmark} & \textbf{PI} & \textbf{PO} & \textbf{LUTs} & \textbf{Cycles} \\ \midrule 
ac97\_ctrl & 2255 & 2250 & 3926 & 27742 \\
comp & 279 & 193 & 8090 & 74379 \\
des\_area & 368 & 72 & 1797 & 17273 \\
div16 & 32 & 32 & 2293 & 22047 \\
hamming & 200 & 7 & 725 & 9414 \\
i2c & 147 & 142 & 423 & 3133 \\
MAC32 & 96 & 65 & 3310 & 40007 \\
max & 512 & 130 & 1866 & 16072 \\
mem\_ctrl & 1198 & 1225 & 3031 & 22021 \\
MUL32 & 64 & 64 & 2758 & 31344 \\
pci\_bridge32 & 3519 & 3528 & 23257 & 110318 \\
pci\_spoci\_ctrl & 85 & 76 & 446 & 3621 \\
revx & 20 & 25 & 3056 & 31603 \\
sasc & 133 & 132 & 204 & 1476 \\
simple\_spi & 148 & 147 & 305 & 2307 \\
spi & 274 & 276 & 1581 & 13115 \\
sqrt32 & 32 & 16 & 989 & 11326 \\
square & 64 & 127 & 6083 & 67602 \\
ss\_pcm & 106 & 98 & 159 & 968 \\
systemcaes & 930 & 819 & 3207 & 26981 \\
systemcdes & 314 & 258 & 1128 & 9468 \\
tv80 & 373 & 404 & 3044 & 25986 \\
usb\_funct & 1860 & 1846 & 5265 & 41029 \\
usb\_phy & 113 & 111 & 187 & 1156 \\ \toprule
\end{tabular}
\end{table}



    

\begin{figure*}
\centering
\begin{minipage}[b][][b]{.60\textwidth}
  \centering
   \begin{subfigure}[t]{0.45\columnwidth}
    {\scriptsize
    \centering 
    \caption{} 
    \label{fig:spacing_tab}
    \renewcommand{\arraystretch}{1.3}
    \setlength{\tabcolsep}{2pt}
    \begin{center}
    \begin{tabular}{cccccccc} \bottomrule 
          &\textbf{Spacing} & \textbf{0} & \textbf{2} & \textbf{4} & \textbf{6} & \textbf{8} \\ \midrule
 $64\times64$& c3540 & 4006 & 3760 & 3702 & 3761 & 3813 \\
 & c5315 & 5354 & 4952 & 4963 & 5032 & 5108 \\
 & c7552 & 8009 & 7348 & 7187 & 7327 & $\times \times$\\ \midrule
  \renewcommand{\arraystretch}{1.4}
 $128\times128$ & c3540 & 3814 & 3664 & 3639 & 3585 & 3614 \\
 & c5315 & 5071 & 4836 & 4795 & 4828 & 4804 \\
 & c7552 & 7807 & 7141 & 7052 & 7038 & 7035 \\ \toprule 
    \end{tabular}
        \end{center}
}
    \end{subfigure}
    \begin{subfigure}[t]{0.50\columnwidth}
    \centering
    \caption{} 
    \label{fig:spacing_viz}
    \vspace{0.2cm}
    \includegraphics[width=\columnwidth]{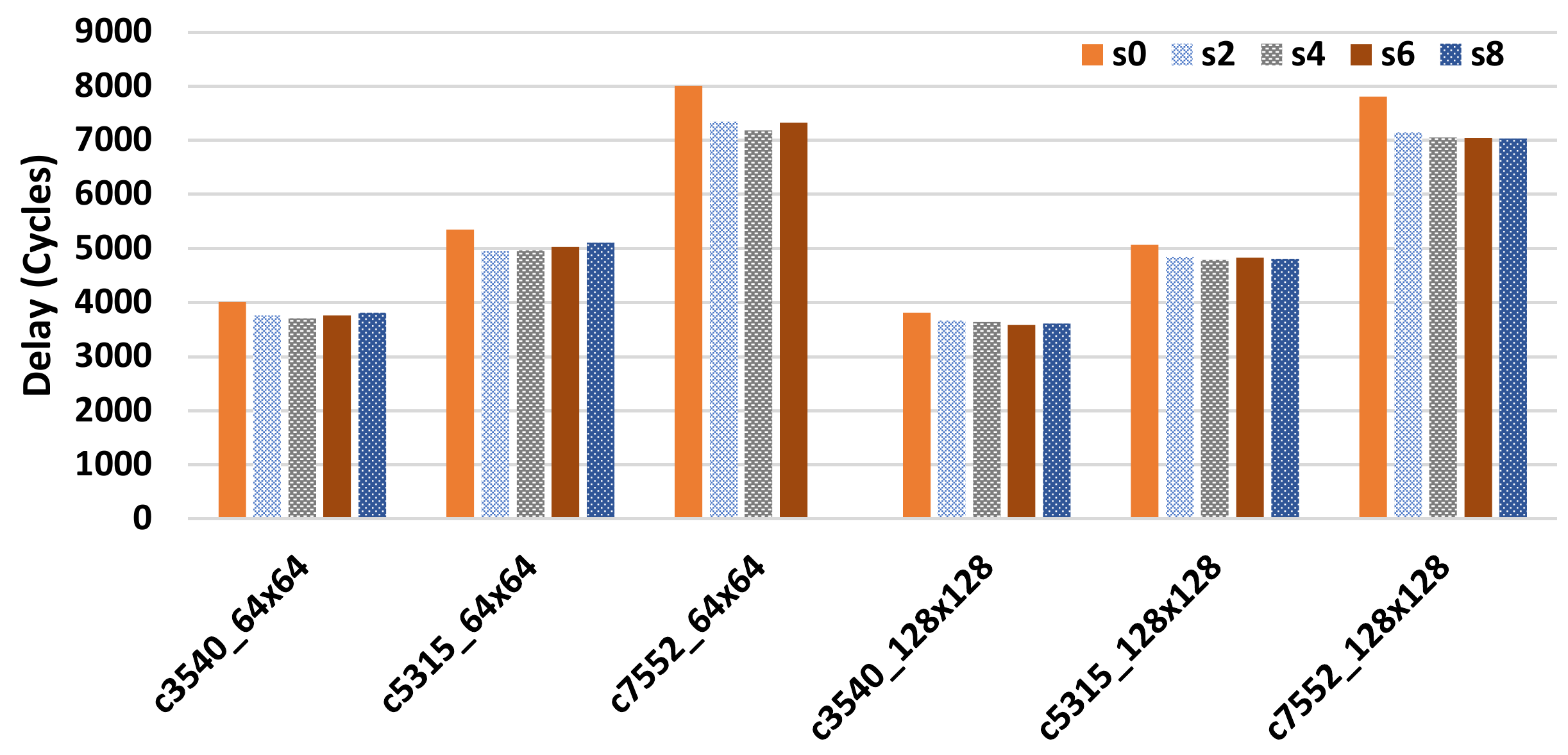}
    \end{subfigure}
  \captionof{figure}{Impact of spacing parameters on delay for three benchmarks, considering two crossbar dimensions $64\times64$ and $128\times128$, with $k=4$.}
    \label{fig:spacing}
\end{minipage}%
\begin{minipage}[b][][b]{.05\textwidth}
	
\end{minipage}
\begin{minipage}[b][][b]{.30\textwidth}
  \centering
  \includegraphics[width=\columnwidth]{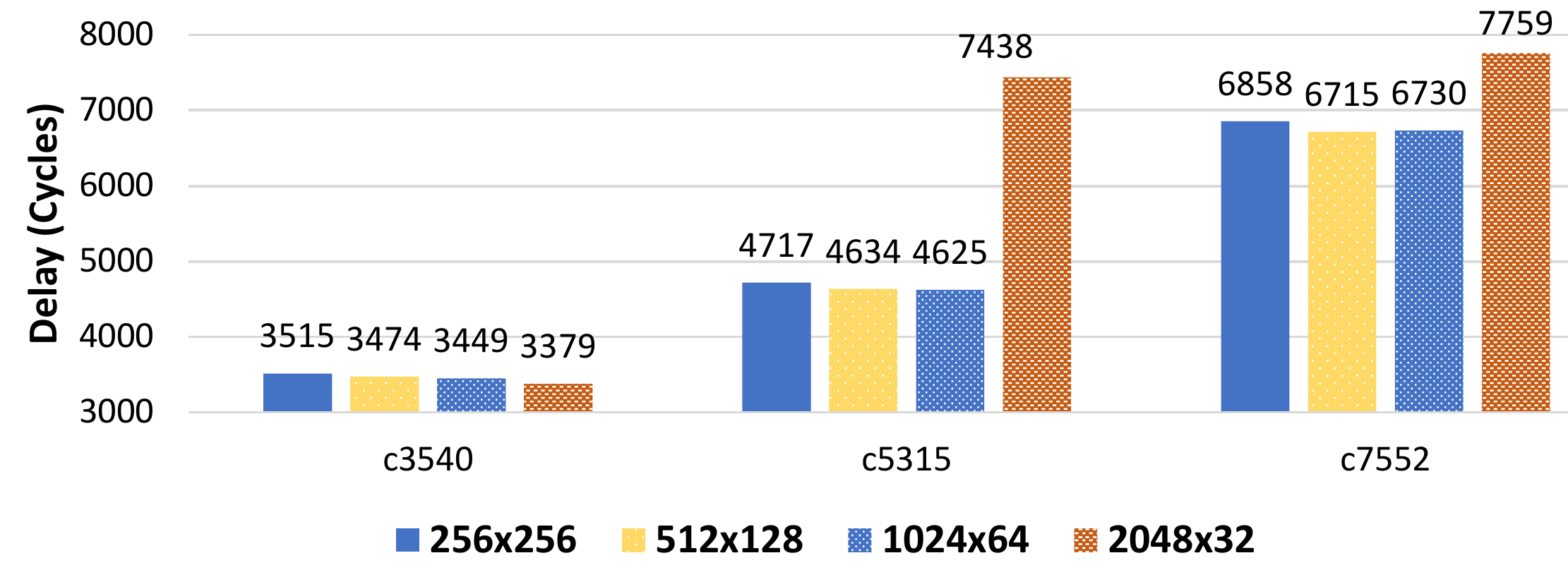}
  \captionof{figure}{Impact of crossbar dimensions on delay of mapping, while keeping the number of devices constant. }
	\label{fig:constArea}
\end{minipage}
\vspace{-0.3cm}
\end{figure*}

\subsection{Impact of spacing parameter} 
{\em Spacing} is the number of rows that is left free between two LUTs stacked vertically, as described in Algorithm~\ref{algo:lutplacement}. We analyze the impact of {\em spacing} on three large benchmarks for ISCAS85, $k=4$ and two crossbar dimensions~$64\times64$ and $128\times128$. The results of analysis are summarily shown in Fig.~\ref{fig:spacing}.  For most of the benchmarks, the delay decreases considerably by increasing spacing from 0 to 4 or 6 (depending on the benchmark). However, increasing spacing further leads to increase in delay. This is due to the fact that leaving empty row helps in finding shorter paths between source and destination locations on the crossbar while using A* search, that leads to reduction in delay. However, setting a large value~(such as 8 or higher) for the spacing parameter leads to lesser space available in the crossbar for actual placement of the LUTs, which leads to reduction in number of parallel operations and higher delay.


\subsection{Impact of crossbar dimensions} 
Fig.~\ref{fig:constArea} shows the impact of crossbar dimensions on delay of mapping, while keeping the number of devices~($R\times C$) constant. We considered $k=\{2,3,4\}$, $spacing=\{0,2,4,6\}$ and three large benchmarks for ISCAS85 benchmarks. The best delay for all the benchmarks were obtained for $k=4$ and spacing=$6$. We can observe that increasing the number of rows and decreasing the number of columns, the delay of mapping decreases. As discussed in Section~\ref{sec:method}, LUTs are stacked in vertical orientation and can be executed in parallel as long as there are no data dependencies and the number of inputs are same. Increasing the number of rows allows greater number of parallel operations to be executed. When a small number of columns are available, the mapping delay increases (as observed by changing crossbar dimensions from $1024\times64$ to $2048\times32$). This is because lower number of devices are available when columns are blocked during for preserving intermediate results and the alignment overhead increases as well. 

\subsection{Copy overhead}
\begin{figure}

    \setlength{\tabcolsep}{3pt}
    
    \centering 
    \includegraphics[width=0.8\columnwidth]{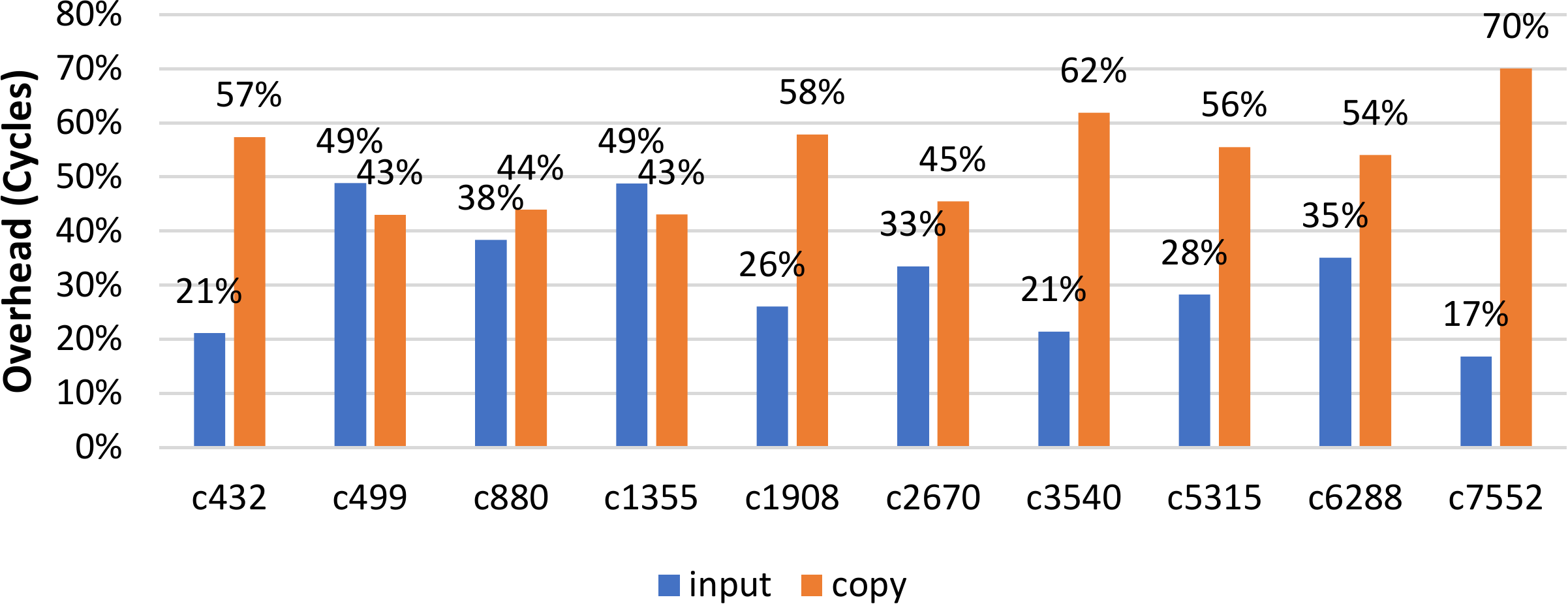}
    
   \caption{Overhead of primary input placement and copying intermediate results for LUT input.} 
    \label{fig:overhead_cyc}

\end{figure}

Fig.~\ref{fig:overhead_cyc} shows the overhead of copy operations as a percentage.
As evident from the Fig.~\ref{fig:overhead_cyc}, copy operations constitute a large overhead in the computation of a benchmark. As we use A* search algorithm to align the inputs, the exact number of copy operations used in alignment is optimal. However in order to limit run time, we do not try and  scheduling multiple copy operations in parallel, considering multiple source and destination locations simultaneously. This could be investigated in future, at the cost of higher execution time of the search algorithm.

\begin{table*}[ht]
    \centering
       \caption{Comparison of CONTRA with existing works. Note that the delay for the existing works do not consider placement overhead of primary inputs. {\small R = Number of Rows, C = Number of Columns, k = Number of inputs to generate LUT Graph.}}
    \label{table:comparison}
        \setlength{\tabcolsep}{1pt}
    \scriptsize
    \begin{tabular}{r|rrr|rr|rr|rr|rr|rr|rr|rrr}\bottomrule
    & \multicolumn{3}{|c|}{Proposed} & \multicolumn{2}{|c|}{E1~\cite{gharpinde2017scalable}} & \multicolumn{2}{c|}{E2~\cite{zulehner2019staircase}} & \multicolumn{2}{c}{E3~\cite{thangkhiew2018scalable}} & \multicolumn{2}{c|}{E4~\cite{thangkhiew2018scalable}} & \multicolumn{2}{c}{E5~\cite{yadav2019look}} & \multicolumn{2}{c|}{E6~\cite{yadav2019look}} & \multicolumn{2}{c}{E7~\cite{tenace2019said}}\\
         \textbf{~~~~~~~~~~~Bench}  & \textbf{k} & \multicolumn{1}{c|}{\textbf{RxC}}  & \textbf{Cycles} & \multicolumn{1}{c|}{\textbf{RxC}}  & \textbf{Cycles} & \multicolumn{1}{c|}{\textbf{RxC}}  & \textbf{Cycles}& \multicolumn{1}{c|}{\textbf{RxC}}  & \textbf{Cycles}& \multicolumn{1}{c|}{\textbf{RxC}}  & \textbf{Cycles}& \multicolumn{1}{c|}{\textbf{RxC}}  & \textbf{Cycles}& \multicolumn{1}{c|}{\textbf{RxC}}  & \textbf{Cycles}& \multicolumn{1}{c|}{\textbf{RxC}}  & \textbf{Cycles} \\ \midrule 
c432  & 3 & 20x12  & 824  & 146x9  & 349  & 22x42  & 225  & 62x11  & 265  & 51x47  & 342  & 36x150  & 338  & 69x13  & 290  & 36x84  & 156  & \\
c499  & 3 & 20x16  & 1140  & 323x13  & 1155  & 96x44  & 242  & 73x37  & 935  & 83x55  & 1059  & 45x182  & 903  & 116x31  & 707  & 144x28  & 420  & \\
c880  & 3 &  32x22  & 1389  & 383x5  & 761  & 67x39  & 427  & 124x30  & 750  & 103x73  & 913  & 69x73  & 726  & 107x14  & 613  & 100x53  & 482  & \\
c1355  & 3 &  36x16  & 1092  & 359x10  & 1072  & 96x63  & 236  & 72x43  & 938  & 91x55  & 1060  & 49x163  & 825  & 103x28  & 757  & 128x37  & 554  & \\
c1908  & 3 &  32x22  & 1489  & 312x13  & 1056  & 83x85  & 517  & 60x60  & 970  & 70x66  & 1075  & 42x88  & 928  & 93x33  & 648  & 69x54  & 627  & \\
c2670  & 4 &  38x34  & 2267  & 664x9  & 1490  & 66x92  & 551  & 301x45  & 1401  & 385x245  & 1495  & 202x137  & 1278  & 340x29  & 1183  & 355x33  & 643  & \\
c3540  & 4 &  60x26  & 3726  & 650x16  & 2396  & 137x164  & 1435  & 153x150  & 2418  & 160x161  & 2589  & 71x221  & 2007  & 109x55  & 1761  & 234x77  & 1566  & \\
c5315  & 4 &  64x48  & 5365  & 1261x11  & 3295  & 221x136  & 1361  & 298x73  & 3239  & 449x179  & 3382  & 249x122  & 2676  & 547x22  & 2251  & 441x42  & 1754  & \\
c6288  & 4 &  32x30  & 8744  & 2297x6  & 3776  & 151x870  & 3751  & 436x98  & 5007  & 265x265  & 5515  & 33x892  & 3161  & 49x115  & 3104  & 510x226  & 4069  & \\
c7552  & 4 &  64x48  & 8009  & 845x14  & 3929  & 214x175  & 2182  & 321x320  & 3824  & 381x379  & 4012  & 220x57  & 3031  & 542x22  & 2486  & 416x79  & 2565  & \\ \toprule
\multicolumn{4}{c}{\textbf{GeoMean Reduction (Area):}} & 5.9$\times$  &   & 10.8$\times$  &    & 9.4$\times$ &   & 19.8$\times$ &  & 12.3$\times$  & & 4.7$\times$ & & 11.5$\times$\\
\multicolumn{4}{c}{\textbf{GeoMean Overhead (Delay):}} &  & 1.6$\times$  &   & 3.5$\times$  &   & 1.7$\times$  &   & 1.5$\times$  &   & 1.9$\times$ & & 2.2$\times$ & & 2.8$\times$\\
\toprule 

    \end{tabular}
  \vspace{-0.5cm}
\end{table*}

\begin{figure}[ht]
\centering 
    \scriptsize
\includegraphics[width=0.5\textwidth]{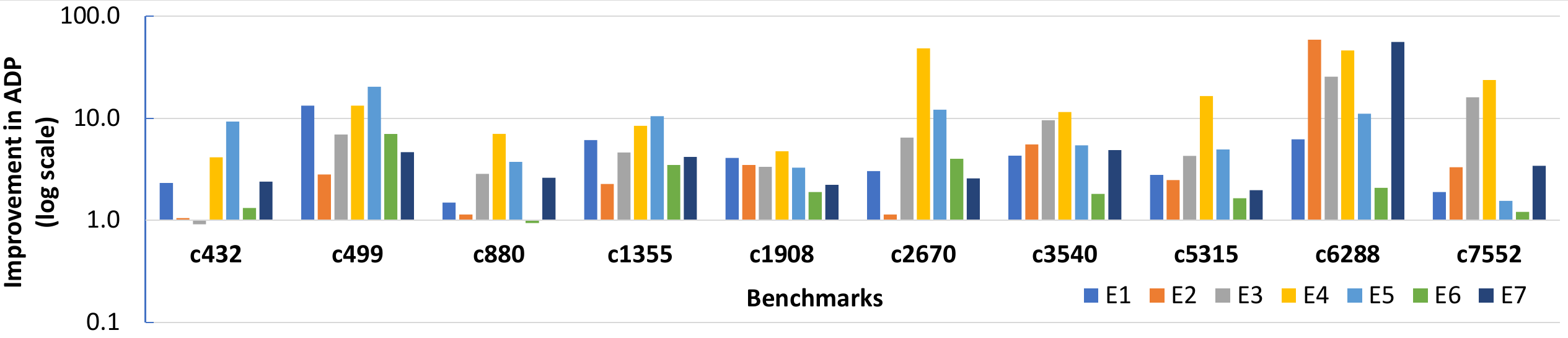}
        \setlength{\tabcolsep}{1pt}
\begin{tabular}{rrrrrrrrrrrrrrrrrrrr}\bottomrule
\multicolumn{4}{c}{\textbf{GeoMean Improvement (ADP):}} &  \multicolumn{2}{c|}{E1: 3.7}  &  \multicolumn{2}{c|}{E2: 3.1} & \multicolumn{2}{c|}{E3: 5.6} &\multicolumn{2}{c|}{E4: 13.1} &\multicolumn{2}{c|}{E5: 6.5} &\multicolumn{2}{c|}{E6: 2.1} &\multicolumn{2}{c|}{E7: 4.1} \\\toprule
\end{tabular}
\caption{Comparison of the ADP of CONTRA with existing works, along with Geometric Mean~(GeoMean) of improvement in ADP of CONTRA over existing works.}
\label{fig:adp}
\end{figure}

\subsection{Comparison with existing works}
The existing technology mapping approaches for MAGIC do not consider area constraints in mapping and focus only on minimizing the delay. Given a benchmark, the existing methods report the crossbar dimensions required to map the benchmark, along with the delay of mapping. These works therefore cannot map benchmarks to arbitrary sized crossbar arrays. For comparison, we determine the smallest crossbar dimension for which the mapping was feasible using CONTRA. In the absence of area constraints, our method can achieve delay identical to SAID (E7)~\cite{tenace2019said}, since both CONTRA and SAID use LUT based mapping.
CONTRA requires significant lower area to map in comparison to existing methods, while having relatively higher delay. As none of the methods support area constraints, we use Area-Delay Product~(ADP) as a composite metric for direct comparison.
 \begin{align}
 ADP &= R \times C \times Cycles\\
 \text{Improvement in } ADP &= \frac{ADP_{Ei}}{ADP_{CONTRA}}
 \end{align}
The list of existing works we compare CONTRA to follows:
\begin{itemize}
    \item $E1$~\cite{gharpinde2017scalable}: A NOR/INV netlist is mapped using MAGIC operations by replicating specific logic levels or gates in order to achieve the maximum parallelism while guaranteeing a square shape allocation of memristors.
    \item $E2$~\cite{zulehner2019staircase}:  A staircase structure is utilized to reach a almost square-like layout with focus on minimizing the number of time steps and utilized memristors.
    \item $E3,E4$~\cite{thangkhiew2018scalable}: These methods correspond to the delay optimisation and crossbar orientation optimisation methods using a simulated annealing approach.
    \item $E5, E6$~\cite{yadav2019look}: These methods correspond to the Look Ahead with Parallel Mapping and  Look Ahead Heuristic and Parallel Mapping methods presented by Yadav et al.  The look-ahead heuristics attempts to minimize the number of copy operations. The parallel mapping approach of the gates tries to maximize the evaluation of gates in parallel. 
    \item $E7$~\cite{tenace2019said}: This method presents a library-free supergate-aided (SAID) logic synthesis approach with a dedicated mapping strategy tailored on MAGIC crossbars relying on LUT-based synthesis. Two main differences exist between this work and the proposed work. Firstly, our proposed approach takes area-constraints as input, where as SAID does not support area constraint. Secondly, our approach does not enforce placement patterns of LUTs which SAID does. Our approach will work with a variety of placement patterns for the LUTs, as the A* search technique can be used for optimally moving the intermediate results to any desired location.
 \end{itemize} 

We present the comparison results in Table~\ref{table:comparison}. The main observations are (1) CONTRA requires less crossbar area compared to all other methods. (2) Not only the total area is smaller, but the size of each dimension is smaller which makes mapping of logic into memory significantly more feasible. (3) Unfortunately, these benefits come with a slightly higher delay. None of the previous works on technology mapping for MAGIC consider the overhead of placing the primary inputs on the crossbar~\cite{gharpinde2017scalable,zulehner2019staircase, thangkhiew2018scalable, tenace2019said, yadav2019look}. However, we considered the cost of  placing the primary inputs in all our mapping results. From Fig.~\ref{fig:overhead_cyc}, we can observe that the overhead of input in terms of number of cycles could be as high as 49\% for smaller benchmarks. This strongly suggests that the overhead of input placement must be considered during mapping. Therefore, comparing our proposed method directly in terms of delay with existing works is unfair.

 In Fig.~\ref{fig:adp}, we plot the improvement in ADP for individual test cases from the ISCAS85 benchmarks. Barring two cases (c432 for E2 and c880 for E6), there is a considerable improvement in ADP for the proposed algorithm for all the benchmarks against all the existing implementations.  We present the geometric mean of improvement in ADP of CONTRA over the  existing methods. CONTRA achieves the best geometric mean improvement of $13.1\times$ over E4. From the Fig.~\ref{fig:adp}, we can also rank existing methods on the basis of their ADP. After CONTRA, E6 has the next best ADP, followed closely by E1 and E2, followed by E7, whereas E3, E4 and E5 are significantly worse.

\subsection{Discussion about Majority based in-memory computing}
Unlike MAGIC operations where all the inputs are represented as state of memristors, Majority operations also use the bitline and wordline inputs as inputs, alongside the internal resistive state $Z$ of the ReRAM which acts as third input and the stored bit.
Using majority operations, ReVAMP architecture was proposed by Bhattacharjee et al.~\cite{bhattacharjee2017revamp}. ReVAMP supports two type of instructions. Apply instructions compute on the cells of a wordline. Read instruction reads the internal state of a word onto a data-memory register by using sense amplifiers, that can be used as input to subsequent Apply instructions. In case of MAGIC, read operations are not used during in-memory operations. 

For the sake of completeness, we compare CONTRA against a recently proposed area-constrained mapping approach~ArC for ReVAMP~\cite{bhattacharjee2018crossbar}. The results of comparison are shown in Table~\ref{table:revamp_comp}. CONTRA achieves better delay compared to ArC, whereas requiring larger number of memristors to map the benchmarks. It should be noted that the delay for ArC is equal to the number of cycles required for computes and reads. Also, ReVAMP uses an external internconnect network for alignment of inputs, which does not contribute to the number of cycles but in practice would imply higher controller energy. In case of MAGIC, alignment operations are done inside the crossbar itself, which leads to higher delay and more number of memristors being used for the COPY operations.

\begin{table}[t]
    \centering
        \caption{Comparison of CONTRA against ArC for \mbox{ReVAMP}~\cite{bhattacharjee2018crossbar}.}
    \label{table:revamp_comp}
    \scriptsize
    \begin{tabular}{r|rc|rc|cr}\bottomrule 
\textbf{Bench} & \textbf{RxC} & \textbf{Overhead} & \textbf{Cycles} & \textbf{Speedup} & $\frac{ADP_{curr}}{ADP_{ArC}}$ & \\ \midrule 
c432 & 8x14 & 2.1 & 1654 & 2.0 & 1.1 & \\
c499 & 8x14 & 2.9 & 2450 & 2.1 & 1.3 & \\
c880 & 8x14 & 6.3 & 2569 & 1.8 & 3.4 & \\
c1355 & 8x14 & 5.1 & 2460 & 2.3 & 2.3 & \\
c1908 & 12x14 & 4.2 & 2774 & 1.9 & 2.2 & \\
c2670 & 16x16 & 5.0 & 4307 & 1.9 & 2.7 & \\
c3540 & 18x24 & 3.6 & 7152 & 1.9 & 1.9 & \\
c5315 & 26x24 & 4.9 & 8005 & 1.5 & 3.3 & \\
c6288 & 16x24 & 2.5 & 14871 & 1.7 & 1.5 & \\
c7552 & 20x24 & 6.4 & 11079 & 1.4 & 4.6 & \\\toprule 
    \end{tabular}
\end{table}
\section{Conclusion}\label{sec:conc}
In this work, we presented the first area-constrained technology mapping flow for LiM using MAGIC operation on a crossbar array. We provide a scalable approach to solve the problem that tries to maximize parallelism. We introduce an optimal search algorithm for alignment of variables between two locations in a crossbar. We unlock the possibility of mapping Boolean functions to a wide variety of crossbar dimensions using MAGIC operations. The proposed algorithm outperforms state-of-the-art technology approaches for MAGIC in terms of ADP.
Evidently from our comparative studies, existing design automation flows for in-memory computing platforms are far from capturing the nuances of practical constraints. To alleviate this problem, we will apply our flow on actual design prototypes and come up with more rigorous benchmarks with detailed characterization. 

{
 
\bibliographystyle{ACM-Reference-Format}
\bibliography{refs}


\begin{thebibliography}{31}


\ifx \showCODEN    \undefined \def \showCODEN     #1{\unskip}     \fi
\ifx \showDOI      \undefined \def \showDOI       #1{#1}\fi
\ifx \showISBNx    \undefined \def \showISBNx     #1{\unskip}     \fi
\ifx \showISBNxiii \undefined \def \showISBNxiii  #1{\unskip}     \fi
\ifx \showISSN     \undefined \def \showISSN      #1{\unskip}     \fi
\ifx \showLCCN     \undefined \def \showLCCN      #1{\unskip}     \fi
\ifx \shownote     \undefined \def \shownote      #1{#1}          \fi
\ifx \showarticletitle \undefined \def \showarticletitle #1{#1}   \fi
\ifx \showURL      \undefined \def \showURL       {\relax}        \fi
\providecommand\bibfield[2]{#2}
\providecommand\bibinfo[2]{#2}
\providecommand\natexlab[1]{#1}
\providecommand\showeprint[2][]{arXiv:#2}

\bibitem[\protect\citeauthoryear{Aga, Jeloka, Subramaniyan, Narayanasamy,
  Blaauw, and Das}{Aga et~al\mbox{.}}{2017}]%
        {aga2017compute}
\bibfield{author}{\bibinfo{person}{Shaizeen Aga}, \bibinfo{person}{Supreet
  Jeloka}, \bibinfo{person}{Arun Subramaniyan}, \bibinfo{person}{Satish
  Narayanasamy}, \bibinfo{person}{David Blaauw}, {and}
  \bibinfo{person}{Reetuparna Das}.} \bibinfo{year}{2017}\natexlab{}.
\newblock \showarticletitle{Compute caches}. In \bibinfo{booktitle}{\emph{2017
  IEEE International Symposium on High Performance Computer Architecture
  (HPCA)}}. IEEE, \bibinfo{pages}{481--492}.
\newblock


\bibitem[\protect\citeauthoryear{Agrawal, Jaiswal, Lee, and Roy}{Agrawal
  et~al\mbox{.}}{2018}]%
        {agrawal2018x}
\bibfield{author}{\bibinfo{person}{Amogh Agrawal}, \bibinfo{person}{Akhilesh
  Jaiswal}, \bibinfo{person}{Chankyu Lee}, {and} \bibinfo{person}{Kaushik
  Roy}.} \bibinfo{year}{2018}\natexlab{}.
\newblock \showarticletitle{X-sram: Enabling in-memory boolean computations in
  cmos static random access memories}.
\newblock \bibinfo{journal}{\emph{IEEE Transactions on Circuits and Systems I:
  Regular Papers}} \bibinfo{volume}{65}, \bibinfo{number}{12}
  (\bibinfo{year}{2018}), \bibinfo{pages}{4219--4232}.
\newblock


\bibitem[\protect\citeauthoryear{Ben-Hur, Ronen, Haj-Ali, Bhattacharjee,
  Eliahu, Peled, and Kvatinsky}{Ben-Hur et~al\mbox{.}}{2019}]%
        {ben2019simpler}
\bibfield{author}{\bibinfo{person}{Rotem Ben-Hur}, \bibinfo{person}{Ronny
  Ronen}, \bibinfo{person}{Ameer Haj-Ali}, \bibinfo{person}{Debjyoti
  Bhattacharjee}, \bibinfo{person}{Adi Eliahu}, \bibinfo{person}{Natan Peled},
  {and} \bibinfo{person}{Shahar Kvatinsky}.} \bibinfo{year}{2019}\natexlab{}.
\newblock \showarticletitle{SIMPLER MAGIC: Synthesis and Mapping of In-Memory
  Logic Executed in a Single Row to Improve Throughput}.
\newblock \bibinfo{journal}{\emph{IEEE Transactions on Computer-Aided Design of
  Integrated Circuits and Systems}} (\bibinfo{year}{2019}).
\newblock


\bibitem[\protect\citeauthoryear{Bhattacharjee, Ama{\'r}u, and
  Chattopadhyay}{Bhattacharjee et~al\mbox{.}}{2018}]%
        {bhattacharjee2018technology}
\bibfield{author}{\bibinfo{person}{Debjyoti Bhattacharjee},
  \bibinfo{person}{Luca Ama{\'r}u}, {and} \bibinfo{person}{Anupam
  Chattopadhyay}.} \bibinfo{year}{2018}\natexlab{}.
\newblock \showarticletitle{{Technology-aware logic synthesis for ReRAM based
  in-memory computing}}. In \bibinfo{booktitle}{\emph{DATE}}.
  \bibinfo{pages}{1435--1440}.
\newblock


\bibitem[\protect\citeauthoryear{Bhattacharjee, Devadoss, and
  Chattopadhyay}{Bhattacharjee et~al\mbox{.}}{2017a}]%
        {bhattacharjee2017revamp}
\bibfield{author}{\bibinfo{person}{Debjyoti Bhattacharjee},
  \bibinfo{person}{Rajeswari Devadoss}, {and} \bibinfo{person}{Anupam
  Chattopadhyay}.} \bibinfo{year}{2017}\natexlab{a}.
\newblock \showarticletitle{ReVAMP: ReRAM based VLIW architecture for in-memory
  computing}. In \bibinfo{booktitle}{\emph{DATE}}. \bibinfo{pages}{782--787}.
\newblock


\bibitem[\protect\citeauthoryear{Bhattacharjee, Easwaran, and
  Chattopadhyay}{Bhattacharjee et~al\mbox{.}}{2017b}]%
        {aomap}
\bibfield{author}{\bibinfo{person}{Debjyoti Bhattacharjee},
  \bibinfo{person}{Arvind Easwaran}, {and} \bibinfo{person}{Anupam
  Chattopadhyay}.} \bibinfo{year}{2017}\natexlab{b}.
\newblock \showarticletitle{{Area-constrained Technology Mapping for In-Memory
  Computing using ReRAM devices}}. In \bibinfo{booktitle}{\emph{22nd Asia and
  South Pacific Design Automation Conference}}.
\newblock


\bibitem[\protect\citeauthoryear{Bhattacharjee, Tavva, Easwaran, and
  Chattopadhyay}{Bhattacharjee et~al\mbox{.}}{2020}]%
        {bhattacharjee2018crossbar}
\bibfield{author}{\bibinfo{person}{Debjyoti Bhattacharjee},
  \bibinfo{person}{Yaswanth Tavva}, \bibinfo{person}{Arvind Easwaran}, {and}
  \bibinfo{person}{Anupam Chattopadhyay}.} \bibinfo{year}{2020}\natexlab{}.
\newblock \showarticletitle{Crossbar-constrained technology mapping for reram
  based in-memory computing}.
\newblock \bibinfo{journal}{\emph{IEEE Trans. Comput.}} \bibinfo{volume}{69},
  \bibinfo{number}{5} (\bibinfo{year}{2020}), \bibinfo{pages}{734--748}.
\newblock


\bibitem[\protect\citeauthoryear{{E. Linn, R. Rosezin, S. Tappertzhofen, U.
  B\"{o}ttger and R. Waser}}{{E. Linn, R. Rosezin, S. Tappertzhofen, U.
  B\"{o}ttger and R. Waser}}{2012}]%
        {eike_logic}
\bibfield{author}{\bibinfo{person}{{E. Linn, R. Rosezin, S. Tappertzhofen, U.
  B\"{o}ttger and R. Waser}}.} \bibinfo{year}{2012}\natexlab{}.
\newblock \showarticletitle{Beyond von Neumann-logic operations in passive
  crossbar arrays alongside memory operations}.
\newblock \bibinfo{journal}{\emph{Nanotechnology}} \bibinfo{volume}{23},
  \bibinfo{number}{30} (\bibinfo{year}{2012}).
\newblock
\urldef\tempurl%
\url{https://doi.org/10.1088/0957-4484/23/30/305205}
\showDOI{\tempurl}


\bibitem[\protect\citeauthoryear{Gaillardon, Amar{\'u}, Siemon, Linn, Waser,
  Chattopadhyay, and Micheli}{Gaillardon et~al\mbox{.}}{2016}]%
        {Gaillardon}
\bibfield{author}{\bibinfo{person}{P.~E. Gaillardon}, \bibinfo{person}{L.
  Amar{\'u}}, \bibinfo{person}{A. Siemon}, \bibinfo{person}{E. Linn},
  \bibinfo{person}{R. Waser}, \bibinfo{person}{A. Chattopadhyay}, {and}
  \bibinfo{person}{G.~De Micheli}.} \bibinfo{year}{2016}\natexlab{}.
\newblock \showarticletitle{The Programmable Logic-in-Memory (PLiM) computer}.
  In \bibinfo{booktitle}{\emph{DATE}}. \bibinfo{pages}{427--432}.
\newblock


\bibitem[\protect\citeauthoryear{Gharpinde, Thangkhiew, Datta, and
  Sengupta}{Gharpinde et~al\mbox{.}}{2017}]%
        {gharpinde2017scalable}
\bibfield{author}{\bibinfo{person}{Rahul Gharpinde},
  \bibinfo{person}{Phrangboklang~Lynton Thangkhiew}, \bibinfo{person}{Kamalika
  Datta}, {and} \bibinfo{person}{Indranil Sengupta}.}
  \bibinfo{year}{2017}\natexlab{}.
\newblock \showarticletitle{A scalable in-memory logic synthesis approach using
  memristor crossbar}.
\newblock \bibinfo{journal}{\emph{IEEE Transactions on Very Large Scale
  Integration (VLSI) Systems}} \bibinfo{volume}{26}, \bibinfo{number}{2}
  (\bibinfo{year}{2017}), \bibinfo{pages}{355--366}.
\newblock


\bibitem[\protect\citeauthoryear{Haj-Ali, Ben-Hur, Wald, Ronen, and
  Kvatinsky}{Haj-Ali et~al\mbox{.}}{2018}]%
        {haj2018not}
\bibfield{author}{\bibinfo{person}{Ameer Haj-Ali}, \bibinfo{person}{Rotem
  Ben-Hur}, \bibinfo{person}{Nimrod Wald}, \bibinfo{person}{Ronny Ronen}, {and}
  \bibinfo{person}{Shahar Kvatinsky}.} \bibinfo{year}{2018}\natexlab{}.
\newblock \showarticletitle{Not in name alone: A memristive memory processing
  unit for real in-memory processing}.
\newblock \bibinfo{journal}{\emph{IEEE Micro}} \bibinfo{volume}{38},
  \bibinfo{number}{5} (\bibinfo{year}{2018}), \bibinfo{pages}{13--21}.
\newblock


\bibitem[\protect\citeauthoryear{Hamdioui, Xie, Nguyen, Taouil, Bertels,
  Corporaal, Jiao, Catthoor, Wouters, Eike, et~al\mbox{.}}{Hamdioui
  et~al\mbox{.}}{2015}]%
        {hamdioui2015memristor}
\bibfield{author}{\bibinfo{person}{Said Hamdioui}, \bibinfo{person}{Lei Xie},
  \bibinfo{person}{Hoang Anh~Du Nguyen}, \bibinfo{person}{Mottaqiallah Taouil},
  \bibinfo{person}{Koen Bertels}, \bibinfo{person}{Henk Corporaal},
  \bibinfo{person}{Hailong Jiao}, \bibinfo{person}{Francky Catthoor},
  \bibinfo{person}{Dirk Wouters}, \bibinfo{person}{Linn Eike}, {et~al\mbox{.}}}
  \bibinfo{year}{2015}\natexlab{}.
\newblock \showarticletitle{Memristor based computation-in-memory architecture
  for data-intensive applications}. In \bibinfo{booktitle}{\emph{DATE}}. EDA
  Consortium, \bibinfo{pages}{1718--1725}.
\newblock


\bibitem[\protect\citeauthoryear{Hansen, Yalcin, and Hayes}{Hansen
  et~al\mbox{.}}{1999}]%
        {hansen1999unveiling}
\bibfield{author}{\bibinfo{person}{Mark~C Hansen}, \bibinfo{person}{Hakan
  Yalcin}, {and} \bibinfo{person}{John~P Hayes}.}
  \bibinfo{year}{1999}\natexlab{}.
\newblock \showarticletitle{Unveiling the ISCAS-85 benchmarks: A case study in
  reverse engineering}.
\newblock \bibinfo{journal}{\emph{IEEE Design \& Test of Computers}}
  \bibinfo{volume}{16}, \bibinfo{number}{3} (\bibinfo{year}{1999}),
  \bibinfo{pages}{72--80}.
\newblock


\bibitem[\protect\citeauthoryear{Hur, Wald, Talati, and Kvatinsky}{Hur
  et~al\mbox{.}}{2017}]%
        {hursimple}
\bibfield{author}{\bibinfo{person}{Rotem~Ben Hur}, \bibinfo{person}{Nimrod
  Wald}, \bibinfo{person}{Nishil Talati}, {and} \bibinfo{person}{Shahar
  Kvatinsky}.} \bibinfo{year}{2017}\natexlab{}.
\newblock \showarticletitle{{SIMPLE MAGIC: synthesis and in-memory mapping of
  logic execution for memristor-aided logic}}. In
  \bibinfo{booktitle}{\emph{Proceedings of the 36th International Conference on
  Computer-Aided Design}}. \bibinfo{pages}{225--232}.
\newblock


\bibitem[\protect\citeauthoryear{Kingra, Parmar, Chang, Hudec, Hou, and
  Suri}{Kingra et~al\mbox{.}}{2020}]%
        {kingra2020slim}
\bibfield{author}{\bibinfo{person}{Sandeep~Kaur Kingra}, \bibinfo{person}{Vivek
  Parmar}, \bibinfo{person}{Che-Chia Chang}, \bibinfo{person}{Boris Hudec},
  \bibinfo{person}{Tuo-Hung Hou}, {and} \bibinfo{person}{Manan Suri}.}
  \bibinfo{year}{2020}\natexlab{}.
\newblock \showarticletitle{SLIM: Simultaneous Logic-in-Memory Computing
  Exploiting Bilayer Analog OxRAM Devices}.
\newblock \bibinfo{journal}{\emph{Scientific reports}} \bibinfo{volume}{10},
  \bibinfo{number}{1} (\bibinfo{year}{2020}), \bibinfo{pages}{1--14}.
\newblock


\bibitem[\protect\citeauthoryear{Kvatinsky, Belousov, Liman, Satat, Wald,
  Friedman, Kolodny, and Weiser}{Kvatinsky et~al\mbox{.}}{2014}]%
        {kvatinsky2014magic}
\bibfield{author}{\bibinfo{person}{Shahar Kvatinsky}, \bibinfo{person}{Dmitry
  Belousov}, \bibinfo{person}{Slavik Liman}, \bibinfo{person}{Guy Satat},
  \bibinfo{person}{Nimrod Wald}, \bibinfo{person}{Eby~G Friedman},
  \bibinfo{person}{Avinoam Kolodny}, {and} \bibinfo{person}{Uri~C Weiser}.}
  \bibinfo{year}{2014}\natexlab{}.
\newblock \showarticletitle{MAGIC—Memristor-aided logic}.
\newblock \bibinfo{journal}{\emph{IEEE Transactions on Circuits and Systems II:
  Express Briefs}} \bibinfo{volume}{61}, \bibinfo{number}{11}
  (\bibinfo{year}{2014}), \bibinfo{pages}{895--899}.
\newblock


\bibitem[\protect\citeauthoryear{Lee, Lin, Lien, Chih, and Chang}{Lee
  et~al\mbox{.}}{2017}]%
        {lee20171}
\bibfield{author}{\bibinfo{person}{Chia-Fu Lee}, \bibinfo{person}{Hon-Jarn
  Lin}, \bibinfo{person}{Chiu-Wang Lien}, \bibinfo{person}{Yu-Der Chih}, {and}
  \bibinfo{person}{Jonathan Chang}.} \bibinfo{year}{2017}\natexlab{}.
\newblock \showarticletitle{A 1.4 Mb 40-nm embedded ReRAM macro with 0.07 um 2
  bit cell, 2.7 mA/100MHz low-power read and hybrid write verify for high
  endurance application}. In \bibinfo{booktitle}{\emph{2017 IEEE Asian
  Solid-State Circuits Conference (A-SSCC)}}. IEEE, \bibinfo{pages}{9--12}.
\newblock


\bibitem[\protect\citeauthoryear{Lehtonen and Laiho}{Lehtonen and
  Laiho}{2009}]%
        {lehtonen2009stateful}
\bibfield{author}{\bibinfo{person}{Eero Lehtonen} {and} \bibinfo{person}{Mika
  Laiho}.} \bibinfo{year}{2009}\natexlab{}.
\newblock \showarticletitle{Stateful implication logic with memristors}. In
  \bibinfo{booktitle}{\emph{NanoArch}}. IEEE Computer Society,
  \bibinfo{pages}{33--36}.
\newblock


\bibitem[\protect\citeauthoryear{Lehtonen, Poikonen, and Laiho}{Lehtonen
  et~al\mbox{.}}{2010}]%
        {lehtonen2010two}
\bibfield{author}{\bibinfo{person}{Eero Lehtonen}, \bibinfo{person}{JH
  Poikonen}, {and} \bibinfo{person}{Mika Laiho}.}
  \bibinfo{year}{2010}\natexlab{}.
\newblock \showarticletitle{Two memristors suffice to compute all Boolean
  functions}.
\newblock \bibinfo{journal}{\emph{Electronics letters}} \bibinfo{volume}{46},
  \bibinfo{number}{3} (\bibinfo{year}{2010}), \bibinfo{pages}{239--240}.
\newblock


\bibitem[\protect\citeauthoryear{Pedram, Richardson, Horowitz, Galal, and
  Kvatinsky}{Pedram et~al\mbox{.}}{2016}]%
        {pedram2016dark}
\bibfield{author}{\bibinfo{person}{Ardavan Pedram}, \bibinfo{person}{Stephen
  Richardson}, \bibinfo{person}{Mark Horowitz}, \bibinfo{person}{Sameh Galal},
  {and} \bibinfo{person}{Shahar Kvatinsky}.} \bibinfo{year}{2016}\natexlab{}.
\newblock \showarticletitle{Dark memory and accelerator-rich system
  optimization in the dark silicon era}.
\newblock \bibinfo{journal}{\emph{IEEE Design \& Test}} \bibinfo{volume}{34},
  \bibinfo{number}{2} (\bibinfo{year}{2016}), \bibinfo{pages}{39--50}.
\newblock


\bibitem[\protect\citeauthoryear{Reuben, Ben-Hur, Wald, Talati, Ali,
  Gaillardon, and Kvatinsky}{Reuben et~al\mbox{.}}{2017}]%
        {reuben2017memristive}
\bibfield{author}{\bibinfo{person}{John Reuben}, \bibinfo{person}{Rotem
  Ben-Hur}, \bibinfo{person}{Nimrod Wald}, \bibinfo{person}{Nishil Talati},
  \bibinfo{person}{Ameer~Haj Ali}, \bibinfo{person}{Pierre-Emmanuel
  Gaillardon}, {and} \bibinfo{person}{Shahar Kvatinsky}.}
  \bibinfo{year}{2017}\natexlab{}.
\newblock \showarticletitle{Memristive logic: A framework for evaluation and
  comparison}. In \bibinfo{booktitle}{\emph{2017 27th International Symposium
  on Power and Timing Modeling, Optimization and Simulation (PATMOS)}}. IEEE,
  \bibinfo{pages}{1--8}.
\newblock


\bibitem[\protect\citeauthoryear{Seshadri, Lee, Mullins, Hassan, Boroumand,
  Kim, Kozuch, Mutlu, Gibbons, and Mowry}{Seshadri et~al\mbox{.}}{2017}]%
        {seshadri2017ambit}
\bibfield{author}{\bibinfo{person}{Vivek Seshadri}, \bibinfo{person}{Donghyuk
  Lee}, \bibinfo{person}{Thomas Mullins}, \bibinfo{person}{Hasan Hassan},
  \bibinfo{person}{Amirali Boroumand}, \bibinfo{person}{Jeremie Kim},
  \bibinfo{person}{Michael~A Kozuch}, \bibinfo{person}{Onur Mutlu},
  \bibinfo{person}{Phillip~B Gibbons}, {and} \bibinfo{person}{Todd~C Mowry}.}
  \bibinfo{year}{2017}\natexlab{}.
\newblock \showarticletitle{Ambit: In-memory accelerator for bulk bitwise
  operations using commodity DRAM technology}. In
  \bibinfo{booktitle}{\emph{Proceedings of the 50th Annual IEEE/ACM
  International Symposium on Microarchitecture}}. \bibinfo{pages}{273--287}.
\newblock


\bibitem[\protect\citeauthoryear{{Shirinzadeh}, {Soeken}, {Gaillardon}, and
  {Drechsler}}{{Shirinzadeh} et~al\mbox{.}}{2018}]%
        {shirin2018TCAD}
\bibfield{author}{\bibinfo{person}{S. {Shirinzadeh}}, \bibinfo{person}{M.
  {Soeken}}, \bibinfo{person}{P. {Gaillardon}}, {and} \bibinfo{person}{R.
  {Drechsler}}.} \bibinfo{year}{2018}\natexlab{}.
\newblock \showarticletitle{Logic Synthesis for RRAM-Based In-Memory
  Computing}.
\newblock \bibinfo{journal}{\emph{IEEE Transactions on Computer-Aided Design of
  Integrated Circuits and Systems}} \bibinfo{volume}{37}, \bibinfo{number}{7}
  (\bibinfo{year}{2018}), \bibinfo{pages}{1422--1435}.
\newblock


\bibitem[\protect\citeauthoryear{Soeken and Chattopadhyay}{Soeken and
  Chattopadhyay}{2016}]%
        {soeken2016unlocking}
\bibfield{author}{\bibinfo{person}{Mathias Soeken} {and}
  \bibinfo{person}{Anupam Chattopadhyay}.} \bibinfo{year}{2016}\natexlab{}.
\newblock \showarticletitle{Unlocking efficiency and scalability of reversible
  logic synthesis using conventional logic synthesis}. In
  \bibinfo{booktitle}{\emph{2016 53nd ACM/EDAC/IEEE Design Automation
  Conference (DAC)}}. IEEE, \bibinfo{pages}{1--6}.
\newblock


\bibitem[\protect\citeauthoryear{Synthesis and Group}{Synthesis and
  Group}{2016}]%
        {abc}
\bibfield{author}{\bibinfo{person}{Berkeley~Logic Synthesis} {and}
  \bibinfo{person}{Verification Group}.} \bibinfo{year}{2016}\natexlab{}.
\newblock \bibinfo{title}{{ABC: A System for Sequential Synthesis and
  Verification}}.
\newblock
  \bibinfo{howpublished}{\url{http://www.eecs.berkeley.edu/~alanmi/abc/}}.
\newblock


\bibitem[\protect\citeauthoryear{Talati, Gupta, Mane, and Kvatinsky}{Talati
  et~al\mbox{.}}{2016}]%
        {talati2016logic}
\bibfield{author}{\bibinfo{person}{Nishil Talati}, \bibinfo{person}{Saransh
  Gupta}, \bibinfo{person}{Pravin Mane}, {and} \bibinfo{person}{Shahar
  Kvatinsky}.} \bibinfo{year}{2016}\natexlab{}.
\newblock \showarticletitle{Logic design within memristive memories using
  memristor-aided loGIC (MAGIC)}.
\newblock \bibinfo{journal}{\emph{IEEE Transactions on Nanotechnology}}
  \bibinfo{volume}{15}, \bibinfo{number}{4} (\bibinfo{year}{2016}),
  \bibinfo{pages}{635--650}.
\newblock


\bibitem[\protect\citeauthoryear{Tenace, Rizzo, Bhattacharjee, Chattopadhyay,
  and Calimera}{Tenace et~al\mbox{.}}{2019}]%
        {tenace2019said}
\bibfield{author}{\bibinfo{person}{Valerio Tenace}, \bibinfo{person}{Roberto~G
  Rizzo}, \bibinfo{person}{Debjyoti Bhattacharjee}, \bibinfo{person}{Anupam
  Chattopadhyay}, {and} \bibinfo{person}{Andrea Calimera}.}
  \bibinfo{year}{2019}\natexlab{}.
\newblock \showarticletitle{SAID: A Supergate-Aided Logic Synthesis Flow for
  Memristive Crossbars}. In \bibinfo{booktitle}{\emph{2019 Design, Automation
  \& Test in Europe Conference \& Exhibition (DATE)}}. IEEE,
  \bibinfo{pages}{372--377}.
\newblock


\bibitem[\protect\citeauthoryear{Thangkhiew and Datta}{Thangkhiew and
  Datta}{2018}]%
        {thangkhiew2018scalable}
\bibfield{author}{\bibinfo{person}{Phrangboklang~L Thangkhiew} {and}
  \bibinfo{person}{Kamalika Datta}.} \bibinfo{year}{2018}\natexlab{}.
\newblock \showarticletitle{{Scalable in-memory mapping of Boolean functions in
  memristive crossbar array using simulated annealing}}.
\newblock \bibinfo{journal}{\emph{Journal of Systems Architecture}}
  \bibinfo{volume}{89} (\bibinfo{year}{2018}), \bibinfo{pages}{49--59}.
\newblock


\bibitem[\protect\citeauthoryear{Xue, Jian, Yang, Xiao, Chen, Xu, Xie, Lin,
  Huang, Zou, et~al\mbox{.}}{Xue et~al\mbox{.}}{2013}]%
        {xue20130}
\bibfield{author}{\bibinfo{person}{Xiaoyong Xue}, \bibinfo{person}{Wenxiang
  Jian}, \bibinfo{person}{Jianguo Yang}, \bibinfo{person}{Fanjie Xiao},
  \bibinfo{person}{Gang Chen}, \bibinfo{person}{Shuliu Xu},
  \bibinfo{person}{Yufeng Xie}, \bibinfo{person}{Yinyin Lin},
  \bibinfo{person}{Ryan Huang}, \bibinfo{person}{Qingtian Zou},
  {et~al\mbox{.}}} \bibinfo{year}{2013}\natexlab{}.
\newblock \showarticletitle{A 0.13 µm 8 Mb Logic-Based Cu $x$Si$_y$O ReRAM
  With Self-Adaptive Operation for Yield Enhancement and Power Reduction}.
\newblock \bibinfo{journal}{\emph{IEEE Journal of solid-state circuits}}
  \bibinfo{volume}{48}, \bibinfo{number}{5} (\bibinfo{year}{2013}),
  \bibinfo{pages}{1315--1322}.
\newblock


\bibitem[\protect\citeauthoryear{Yadav, Thangkhiew, and Datta}{Yadav
  et~al\mbox{.}}{2019}]%
        {yadav2019look}
\bibfield{author}{\bibinfo{person}{Dev~Narayan Yadav},
  \bibinfo{person}{Phrangboklang~L Thangkhiew}, {and} \bibinfo{person}{Kamalika
  Datta}.} \bibinfo{year}{2019}\natexlab{}.
\newblock \showarticletitle{Look-ahead mapping of Boolean functions in
  memristive crossbar array}.
\newblock \bibinfo{journal}{\emph{Integration}}  \bibinfo{volume}{64}
  (\bibinfo{year}{2019}), \bibinfo{pages}{152--162}.
\newblock


\bibitem[\protect\citeauthoryear{Zulehner, Datta, Sengupta, and Wille}{Zulehner
  et~al\mbox{.}}{2019}]%
        {zulehner2019staircase}
\bibfield{author}{\bibinfo{person}{Alwin Zulehner}, \bibinfo{person}{Kamalika
  Datta}, \bibinfo{person}{Indranil Sengupta}, {and} \bibinfo{person}{Robert
  Wille}.} \bibinfo{year}{2019}\natexlab{}.
\newblock \showarticletitle{A staircase structure for scalable and efficient
  synthesis of memristor-aided logic}. In \bibinfo{booktitle}{\emph{Proceedings
  of the 24th Asia and South Pacific Design Automation Conference}}. ACM,
  \bibinfo{pages}{237--242}.
\newblock


\end{thebibliography}
}
\end{document}